\documentclass{aa}  

\usepackage{graphicx}
\usepackage{multicol, multirow}
\usepackage{xcolor}
\usepackage{tabularx}
 \usepackage[normalem]{ulem}
 \usepackage{amsmath}	% Advanced maths commands
\usepackage{longtable, pdflscape, booktabs}
\usepackage{lipsum}
\usepackage{blindtext}
\usepackage{txfonts}
\usepackage{hyperref}
\hypersetup{
    colorlinks=true,
    linkcolor=black,      
    urlcolor=black,
    citecolor=blue,
    }
\newcommand*\revise[1]{{\textcolor{black}{#1}}}
\newcommand*\afe{[$\alpha$/Fe]}
\newcommand*\teff{$T_{\rm eff}$}
\newcommand*\feh{[Fe/H]}
\newcommand*\logg{$\log g$}

\newcommand*\pgs{\texttt{PyGlobsterS}}
\newcommand*\atlas{\texttt{ATLAS12}}
\newcommand*\synthe{\texttt{SYNTHE}}

\newcommand*\adev{$\widetilde\Delta$}

\newcommand*\CNa{CN$_1$}
\newcommand*\CNb{CN$_2$}
\newcommand*\Ca{Ca4227}
\newcommand*\Gindex{G4300}
\newcommand*\NaD{Na D}

\begin{document} 

   \title{Synthetic stellar spectra to study\\multiple populations in globular clusters: }

   \subtitle{an extended grid and the effects on the integrated light}

   \author{Vinicius Branco\inst{1,2}
          \and
          Paula R. T. Coelho\inst{1}
          % \fnmsep\thanks{Just to show the usage of the elements in the author field}
          \and
          Ariane Lançon\inst{2}
          \and
          Lucimara P. Martins\inst{3}
          \and
          Philippe Prugniel\inst{4}
          }

% email:philippe.prugniel@univ-lyon1.fr

   \institute{
    Universidade de São Paulo, IAG, Rua do Matão, 1226, 05508-090, Sao Paulo, SP, Brazil.\\ \email{vbranco@usp.br}
    \and
     Université de Strasbourg, CNRS, Observatoire astronomique de Strasbourg, UMR 7550, F-67000 Strasbourg, France.
     \and
     NAT – Universidade Cidade de São Paulo, Rua Galvão Bueno, 868, 01506-000, Sao Paulo, SP, Brazil.
     \and
     Université de Lyon, LyonI, CRAL-Observatoire de Lyon UMR5574, CNRS, France. 
     }
     
   \date{Received Dec. 18, 2023; accepted Apr. 16, 2024.}

% \abstract{}{}{}{}{} 
% 5 {} token are mandatory

% AL : I reformatted the abstract, because the subsections are NOT actually mandatory and I absolutely hate them. 
  \abstract{
   Most Galactic Globular Clusters (GCs) harbour multiple populations of stars (MPs), composed of at least two generations: the first characterized by a “standard” $\alpha$-enhanced metal mixture, as observed in field halo stars of the Milky Way, and the second displaying anti-correlated CN--ONa chemical abundance pattern in combination with an enhanced helium fraction.
   Adequate collections of stellar spectra are needed to characterize the effect of such stellar abundance changes on the integrated light of GCs.
   We present a grid of synthetic stellar spectra \revise{covering the atmospheric parameters relevant to old stellar populations at four subsolar metallicities and two abundance patterns, representative of first- and second-generations of stars in GCs. Integrated spectra of populations were computed using our stellar grid and empirical stellar populations, namely, colour-magnitude diagrams from literature for Galactic GCs.} The spectra range from 290 to 1000nm, where we measured the effect on several spectrophotometric indices due to the surface abundance variations attributed to MPs.
   We find non-negligible effects of the MPs on spectroscopic indices sensitive to C, N, Ca, or Na, and on Balmer indices; we also describe how MPs modify specific regions in the near-UV and near-IR that can be measured with narrow or medium photometric passbands. The effects vary with metallicity.
   A number of these changes remain detectable even when accounting for
   the stochastic fluctuations due to the finite nature of the stellar population cluster.}


   \maketitle

%%%%%%%%%%%%%%%% CONTENT %%%%%%%%%%%%%%%%%%%

\section{Introduction}
\label{sec:introduction}

It is currently well-accepted that most Galactic Globular Clusters (GCs) are characterized by multiple populations of stars (MPs). Evidence has been found in color-magnitude diagrams (CMDs) \citep{Piotto2007, Milone2013, Milone2016, Dondoglio2022, DAntona2022}, \revise{and directly via the spectroscopic determination of star-by-star chemical abundances \citep{Wheeler1989, Kraft1994, Kraft1997, Carretta2009, Carretta2010}. }
Chemical variations among stars of the same GC are found to be anti-correlated, with one element being depleted while the other is enhanced, such as carbon and nitrogen, oxygen and sodium, and sometimes magnesium and aluminium \citep{Bragaglia2010, Gratton2012, VandenBerg2022}. For an extensive study about MPs in GCs and star clusters in general we refer the reader to the reviews by \citet{BastianLardo2018}, \citet{Gratton2019} and \citet{Krumholz2019Areview}.

If Galactic clusters are local representatives of GCs in general, extragalactic GCs \revise{(EGCs)} should present the same MP phenomenon. Indeed, GCs outside of the MW have been extensively studied \citep[e.g.,][]{Brodie2006, Schiavon2013, Larsen2014, Nardiello2018, Salaris2019, Larsen2022, DAbrusco2022}, and the multiple stellar populations have been reported, for example, in the LMC and SMC \citep[e.g.,][]{Mucciarelli2009, Dalessandro2016, Niederhofer2017a, Niederhofer2017b, Hollyhead2017, Gilligan2019, Lagioia2019, Milone2020, Saracino2020, Salgado2022}.
While the mechanism behind the formation of MPs is still unknown 
% \LM{[ref, I will search for some]} \vb{[I'm looking for it too]}, 
the observed integrated spectra of GCs are frequently used to test simple stellar population (SSP) models \citep[see,][]{Lee2005,percival09,walcher09,vazd10,tjm, martins19}.
It is not yet clear to what extent using SSP models to represent systems that are not homogeneous in terms of chemical abundances may impact the analysis of extragalactic GCs \citep[e.g.][]{larsen+18}.

\revise{Efforts have also been directed towards searches for signatures of MPs in the integrated light of EGCs. 
For instance, \citet{McWilliamBernstein2008} have determined abundances of several elements for 47 Tuc, finding enhanced Na and Al. 
Studying 31 GCs from M31, \citet{Colucci2009, Colucci2014, Colucci2017} reported correlations of light element abundance ratios with luminosity and velocity dispersion, evidence of Mg, Na, and Al abundance variations amid GC stars, and that Mg, Al (and likely O, Na) measurements of those EGCs resemble the Galactic GCs. 
\citet{Colucci2011} reported that old GCs in LMC display higher abundance variations of the light elements Mg, Al, and Na than younger GCs, while \citet{Schiavon2013, Sakari2016, Sakari2021} report finding that light-element enhancements show positive correlations with EGC mass. Furthermore, studies have measured the chemical abundances \citep{Larsen2022} and metallicities \citep{Sakari2022} of the Local Group and outer halo M31 GCs, respectively.}

\revise{Regarding the modelling of integrated light, }\citet*[][hereafter ``\textbf{C11}'']{coelho11, coelho+12proc} were the first to \revise{predict} how the chemical anticorrelations affect the integrated spectrum of stellar populations. The authors computed integrated stellar populations with both a ``standard'' $\alpha$-enhanced metal mixture \afe\ $\sim 0.4$ and with an anticorrelated CN--ONa chemical abundances pattern (with and without He enhancement). 
%The stellar populations were computed assuming isochrones from BaSTI \citep{BaSTI2006} and tailored synthetic spectra based on atomic opacities from \citet{coelho+05,castelli_hubrig04}, and ATLAS12 and SYNHTE codes \citep{ATLAS1970, synthe1, atlas12, ATLAS12linux}. The synthetic spectra covered the wavelength region from 3500 to 6000\,\AA, with spectral resolution $R=10\,000$. 
\revise{They provide a quantitative estimate of the maximum effect that a second population would have on Lick indices,  for an iron abundance representative of a typical metal-rich galactic GCs (\feh\ $= -0.7$)}. 
Their results indicate that 
%main spectral indices appreciably affected by the anticorrelations are \CNa, \CNb, \Ca, \Gindex, and \NaD. 
\revise{the presence of a 2nd population would increase the equivalent width of some indices (e.g. H$_\gamma$, CN$_1$, CN$_2$, and NaD) and decrease the equivalent width of others (Ca4226, G4300, and Mg$_b$).}
\revise{These changes go in the direction needed to explain discrepancies between models and GCs observations when only $\alpha$-enhancement chemical changes are taken into account in SSP models \citep{yeps-1}}.
The Balmer lines are affected by the second generation of stars when helium enhancement is considered through the change of the turnoff temperature of the underlying isochrone. \revise{This effect would imply that an integrated spectrum could appear up to 2--3\,Gyr younger than the true age of the population}.
\revise{Yet, C11 predictions were limited to only one iron abundance and the wavelength range 3500 to 6000\,\AA. The effects on a wider range of metallicities and observables remain to be explored.}

In the present work, we aim to expand the study performed by C11, by making a grid of synthetic stellar spectra available for a wider range of metallicities and wavelengths, both for a standard $\alpha$-enhanced metal mixture and for a composition characteristic of second populations of globular clusters. The spectra are computed with an optimized combination of existing line lists. They are used to predict the effects of MPs on the integrated spectra of old clusters and spectrophotometric indices, as a function of metallicity. Because globular clusters contain only a finite number of stars, clusters of a given age, composition, and mass can randomly display a range of integrated properties
\citep[e.g.][]{Barbaro1977,Bruzual2002,Fouesneau2010,Popescu2010,daSilvaSLUG2012}.
We examine to what extent the effects of MPs remain detectable in that stochastic context.

This paper is organized as follows: in Section \ref{sec:grid} we present the synthetic stellar grid, and in Section \ref{sec:application} we describe how the integrated stellar population models were built. We simulate the stochastic populations in Section \ref{sec:bootstrap}, and discuss the results of measured properties in Section \ref{sec:discussion_results}. Our concluding remarks are given in Section \ref{sec:conclusions}.
\section{Synthetic stellar spectra with abundances representative of globular cluster stars}
\label{sec:grid}

We computed {a grid of synthetic stellar spectra suitable for modelling} integrated SSPs {with old ages, subsolar metallicities, and chemical abundance patterns relevant to globular clusters}. Here, we describe the codes and ingredients used and the properties of the final grid.

\subsection{Ingredients and codes}

The model atmospheres and the synthetic spectra were computed with the Linux ports of \atlas\  and \synthe\ codes, respectively \citep{ATLAS1970, atlas12, synthe1, synthe2, atlassynthe1, atlassynthe2}. 
These are the same codes adopted in C11 and have been used recently in the study of the integrated data of GCs \citep[][]{Jang2021, Larsen2022}. \revise{For each chemical mixture pattern adopted in this project, we computed both the model atmosphere and the synthetic spectrum.}

Our atmosphere models were computed using the Opacity Sample method under LTE conditions and 1-D plane-parallel geometry. 
The models were calculated assuming %a set of equally physical and computational parameters required by \atlas\ (i.e., 
the microturbulence of 1 km/s,
60 iterations, 72 layers, 
%the Rosseland optical depth parameter of -6.875, 
mixing length parameter of 1.25, and no overshooting.
{We adopted the same convergence criteria as \citet{mezsaros2012} for the model atmospheres: 
the layers are tested to have the difference in flux and the flux derivative errors to be less than 1\% and 10\%, respectively; }
no more than one non-converged layer was accepted between $\log\tau_{\rm Ross} = -5$ and $\log\tau_{\rm Ross} = 1$, where $\tau_{\rm Ross}$ is the Rosseland optical depth.

% In the end, x perc. converged throughout the whole atmosphere structure.}
% For all models, we tried to converge it exhaustively until the ATLAS12 convergence test failed. The non-converged models were reevaluated assuming \citet{mezsaros2012} convergence criteria.}

The synthetic spectra were computed with a sampling resolution of 1\,700\,000, convolved with a Gaussian filter to a spectral resolution of $R = 850\,000$, from 290\, nm to 950\, nm in the air wavelength. 
Molecular opacities were obtained from R. Kurucz, covering the following molecules:
AlH [A-X], 
AlH [B-X], 
AlO, 
C\textsubscript{2} [A-X], 
C\textsubscript{2} [B-A], 
 C\textsubscript{2} [D-A], 
C\textsubscript{2} [E-A], 
CaH, 
CaO, 
CH, 
CN [A-X], 
CN [B-X], 
CN [X-X], 
CO [A-X], 
CO [X-X], 
CrH [A-X], 
FeH [F-X], 
H\textsubscript{2}, 
MgH, 
MgO, 
NaH, 
NH, 
OH, 
SiH, 
SiO [A-X], 
SiO [E-X], 
SiO [X-X], 
TiH, 
TiO, and
VO
% \pc{add here the name of the molecules}
% \vb{do we need to mention explicitly the molecules names here and refer to the appendix where they are listed with references?}
% \lm{ I think you should list them explicitly here.}
\footnote{ \url{http://kurucz.harvard.edu/linelists/linesmol/}
(see Table \ref{ap:tab:molecules} in the Appendix \ref{ap:A} and \citet{mastersthesis} for details.)}.
{We compiled a new atomic opacity list based on three lists }available in the literature, described in section \ref{calibration}. {The chemical patterns of the grid are discussed in section \ref{sec:coverage}}.
%\LM{[I think this next phrase should be removed from here. Makes more sense to cite and explain in the next section] 
%\sout{\revise{As for the solar abundance pattern, discussed in the next section, we evaluated a simple test computed synthetic spectra with different sources and our compiled opacity list to decide which one to use.}}}
\label{revise_solar_pattern_test}

To automatise the process of computing the grid, we developed a \texttt{Python} wrapper 
called ``Python Globular Cluster Synthesizer'' (hereafter, {\pgs}), which combines the \atlas\ and \synthe\ executions and the integration of the SSP spectra (described later in Section \ref{sec:application}). \pgs\ decreases the time consumed to compute a large sample of models by allowing it to run in parallel jobs.

\subsection{Optimisation of the atomic line list}
\label{calibration}

Different opacities can significantly impact the quality of the synthetic spectrum.
\citet{Martins2007}, for example, tested the accuracy of stellar libraries that assume different opacities and codes, 
observing that the library with the best average performance employed an atomic line list calibrated against the spectra of the Sun and Arcturus. 
Recent studies have pointed out the need for accurate opacity lists, not only to describe stars on different evolutionary stages
but also different regions of the wavelength range \citep[e.g.,][]{Martins2014, franchini+18, ariane21}.

Based on the purpose of this work, we focused on compiling a list based on literature sources that were available in the format required by \synthe:
\citet{coelho14} (hereafter, "\textbf{Coelho14}"), 
\citet{gfkurucz}\footnote{Downloaded from \url{http://kurucz.harvard.edu/linelists/gfnew/gfall08oct17.dat} on Aug 8th, 2018.} (hereafter, "\textbf{Kurucz18}"), and an updated version of the list by \citet{castelli_hubrig04}\footnote{Downloaded from \url{http://wwwuser.oats.inaf.it/castelli/linelists.html} on Aug 8th, 2018.} (hereafter, "\textbf{Castelli16}", corresponding to the version of Feb 18th, 2016).

We computed one synthetic solar spectrum for the three atomic lists, keeping the same molecular opacities unchanged.  
In future work, we plan on expanding this study to newer literature on atomic opacities, which were yet to be available when this project started \citep{Larsen2022, Peterson2022}. 

The Sun was chosen as the reference star given its well-defined stellar parameters and high spectral resolution data available in the literature. 
\revise{We considered three determinations for the solar abundances: \citet{gresau98}, \citet{asplund05}, and \citet{asplund09}. 
All solar models adopt \teff\ = 5777\,K, \logg = 4.4377  \citep[][4th. Ed.]{allenAstropQuant2000} and V\textsubscript{turb} = 1.0\,km/s  \citep{Castelli2003}.} 
% \VB{I could not find the source where Castelli/Kurucz chose these solar atm params. The reference of new ODF they used to compute the solar models may be \citet{Castelli2003}.}

The synthetic solar spectra were compared against the solar spectrum obtained from \citet{wallace11}. This is a high-quality solar spectrum obtained with the Fourier transform spectrometer (FTS) at the McMath-Pierce telescope \citep[as described in][]{Brault1985} which covers the wavelength region from $\sim$\,2958--9250\,\AA\ with resolution varying from $R=\lambda \mathbin{/} \Delta\lambda \sim$\,350,000--700,000, distributed in six regions \citep[see Table 1 in ][]{wallace11}. It is a ground-based spectrum corrected for the effect of telluric lines.

We convolved the synthetic spectra to the relevant spectral resolution 
and used the following metric to quantify the differences between the models and the solar spectrum: 

\begin{equation}
\label{eq:ADEV}
\widetilde\Delta (\lambda) = \frac{1}{N}\sum_{\lambda_{1}}^{\lambda_{2}}\left | \frac{ f_{\rm synt}(\lambda_i)-f_{\rm obs}(\lambda_i) }{ f_{\rm obs}(\lambda_i) } \right |
\end{equation}

\noindent
where $N$ is the number of pixels in wavelength interval $\Delta\lambda = \lambda_{2}-\lambda_{1}$ centred at $\lambda$, and $f_{\rm synt}(\lambda_i)$ and $f_{\rm obs}(\lambda_i)$ are the synthetic and observed flux, respectively, at the $i$-th element of wavelength. 

\begin{table*} \centering
\caption{Global performance of each synthetic spectrum computed with different atomic line lists and different solar abundance patterns, according to the metric in Equation \ref{eq:ADEV}. $\Delta\lambda$ corresponds to 2958--9250\,\AA, the wavelength range of the observed solar spectrum.}

\label{tab:adevopaclist}
\begin{tabular}{|c|c|c|c|c|}
\toprule

\textbf{Solar Pattern} &
\textbf{$\widetilde\Delta$ Castelli16} & 
\textbf{$\widetilde\Delta$ Coelho14} & 
\textbf{$\widetilde\Delta$ Kurucz18} &
\textbf{$\widetilde\Delta$ This Work} 
\\ \toprule

\citet{gresau98} & 
\footnotesize 0.9861 &  %  \pm 0.0042
\footnotesize 0.9894 &  %  \pm 0.0044
\footnotesize 0.9895 &  %  \pm 0.0043
\footnotesize 0.9777 %  \pm 0.0041
\\ 
\citet{asplund05} & 
\footnotesize 0.9842 &  %  \pm 0.0042
\footnotesize 0.9884 &  %  \pm 0.0045
\footnotesize 0.9876 &  %  \pm 0.0044
\footnotesize 0.9763 %  \pm 0.0041
\\ 
\citet{asplund09} & 
\footnotesize 0.9856 &  %  \pm 0.0042
\footnotesize 0.9891 &  %  \pm 0.0045
\footnotesize 0.9891 &  %  \pm 0.0044
\footnotesize 0.9771 %  \pm 0.0041

\\ \bottomrule
\end{tabular}
\end{table*}

\begin{figure}
	\includegraphics[width=\columnwidth]{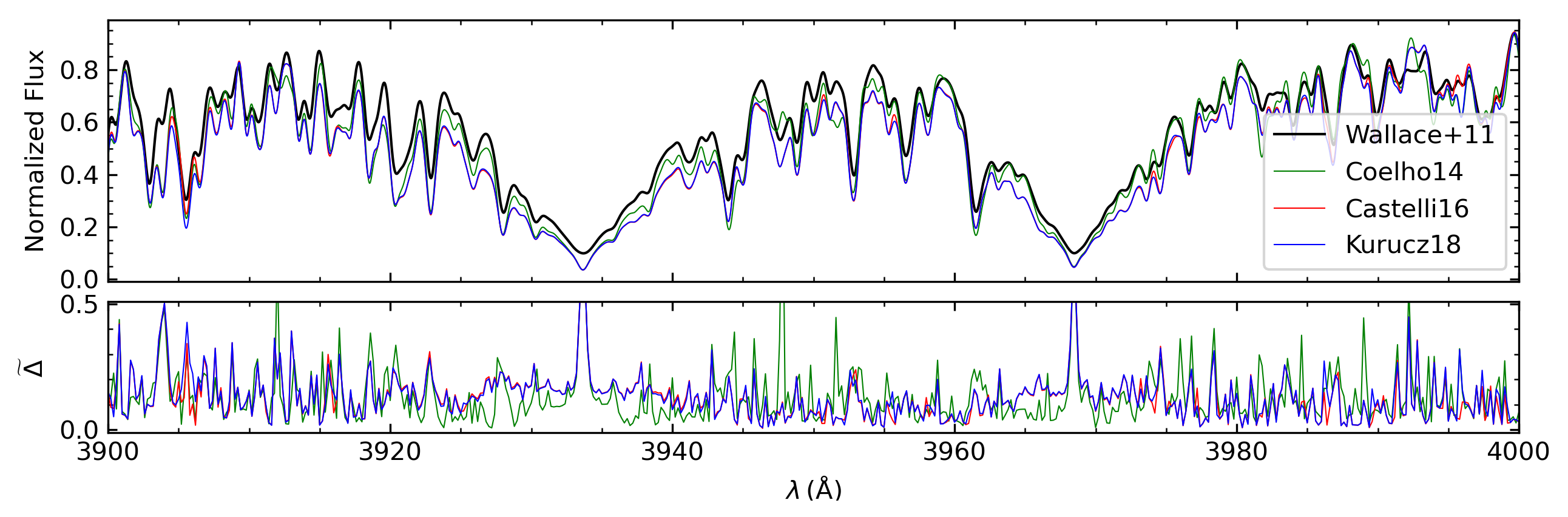}\\
	\includegraphics[width=\columnwidth]{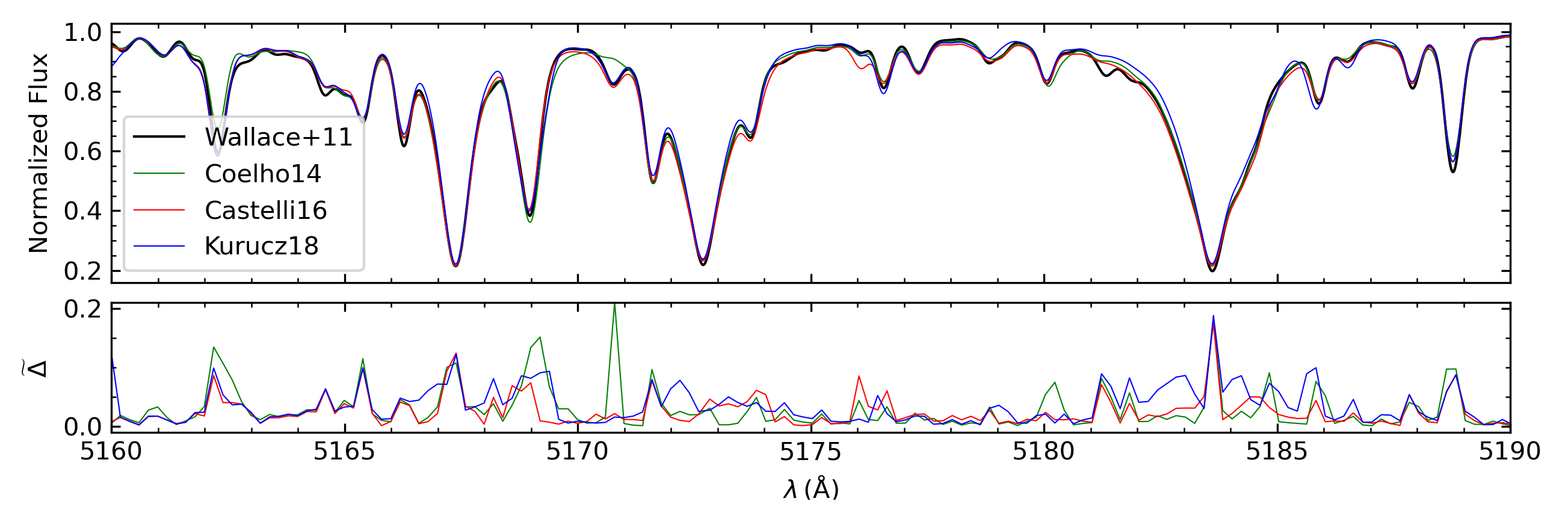}\\
	\includegraphics[width=\columnwidth]{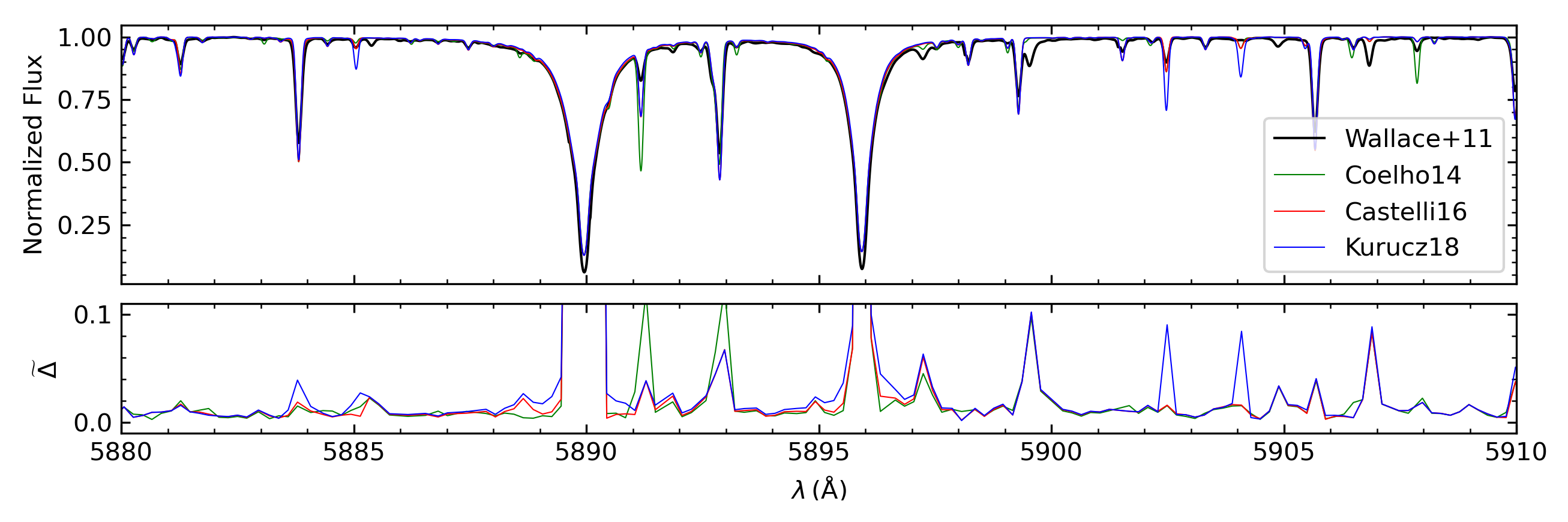}
	\caption{Comparison between modelled and observed spectra (top panels) and their $\widetilde\Delta$ (bottom panels) in three regions: CaII H\&K (top), Mg T (middle), and Na D (bottom). Each panel shows synthetic spectra computed with Coelho14 (green), Castelli16 (red), and Kurucz18 (blue) atomic line lists compared to the observed solar spectrum. $\widetilde\Delta$ is computed according to Eq.\,\ref{eq:ADEV} with $\lambda_2-\lambda_1=0.2\,\AA$. }
    % \caption{\vb{Arrumar legenda} Upper panel: solar observed spectrum (in grey) and the synthetic spectra computed with the three atomic line lists of reference -COE14 (green), CAS16 (blue) and KUR18 (red)- and with the new atomic line list produced in this work (black). Lower panel: performance of each synthetic spectrum compared to the solar spectrum. This plot is arbitrarily degraded for visualization purposes.\pc{Na real não dá pra ver a diferença entre as linhas nessa figura, fica tudo indistinguível e nao dá pra concluir nenhuma message-to-take-home. Acho que vai ter que trabalhar com mais de um painel e zooms. Acho que nao precisamos mostrar todo o intervalo de lambda.} \vb{Colocar o nome dos artigos inteiros ao inves da sigle}}
    \label{fig:adev}
\end{figure}

% \lm{Variations do not seem adequate here. Adev as a funcition of, or Values of adev...}
In Figure \ref{fig:adev} we show the solar spectra {computed with the different line lists} and \adev\ as a function of wavelength, to illustrate the behaviour around CaII H\&K (top), Mg T (middle), and Na D (bottom). 
In general, as the wavelength increases, \adev\ decreases. 
% The synthetic solar spectra were computed with the \citet{asplund09} abundance pattern. 

{To choose between line lists we compared the synthetic spectra with the observed solar spectrum over small wavelength intervals of $\Delta\lambda=0.2$\,\AA}.
For each spectral segment 
%$\Delta \lambda$ in equation (1), we selected
we used equation (1) to select
 the list which best reproduced the solar spectrum (i.e. lowest \adev). The best lists {of each segment} 
 were combined into a new list covering the whole wavelength interval of the observed spectrum.
Table \ref{tab:adevopaclist} summarises the global performance of each line list (i.e. the $\widetilde\Delta$ computed over the wavelength range 2958--9250\,\AA), for three solar abundance patterns. 
{Because the variation among the solar patterns is comparable, we decide to use the most recent reference in the remainder of this article, i.e. \citet{asplund09}.}
\label{revise_choice_solar_pattern_test}

% \lm{what were the abundances used in Figure 1? It is not clear why cite 3 solar abundances. Were the line parameters chosen based on the 3? Is it just for comparison? }
% \vb{Issued addressed.}

\subsection{\label{sec:coverage}Parameter coverage in the Kiel plane}

We aim at a {grid of synthetic} spectra suitable to represent GCs of different subsolar metallicities. The {grid's} coverage of the  Kiel plane (\teff\ vs. \logg) is based on Milky Way GCs selected from \citet[][hereafter "\textbf{M19}"]{martins19}, who have translated empirical CMDs from \citet{Piotto2002} into the \teff\ vs. \logg\ plane using the colour transformations from \citet{worthey_lee11}. 
\revise{The CMDs were observed with the WFPC2 camera of the Hubble Space Telescope. 
The PC was centred in the centre of the clusters and covered a field-of-view of about 2.5' x 2.5'. }

\begin{figure}
	\includegraphics[width=\columnwidth]{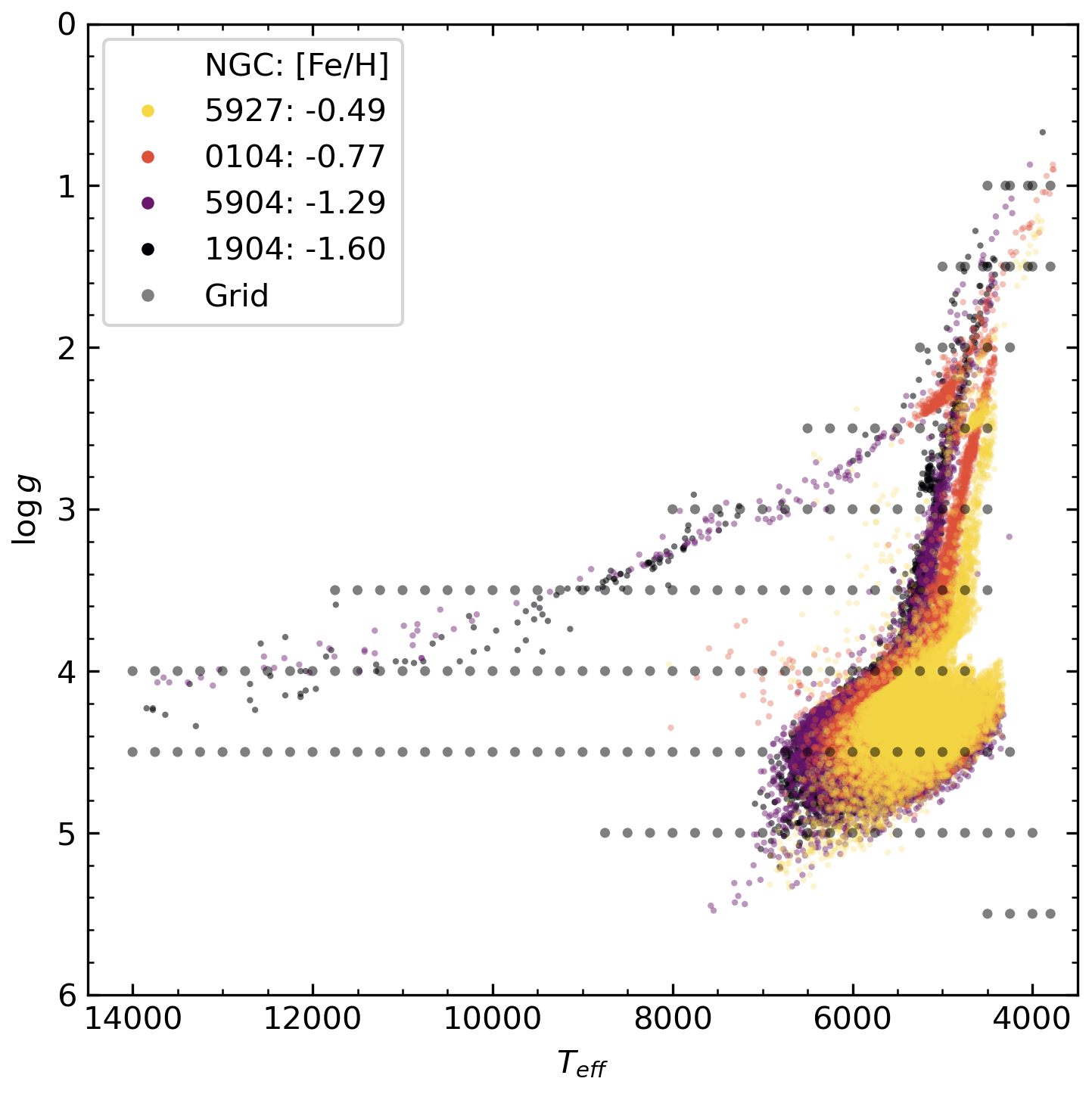}
    \caption{HRD estimated by \citet{martins19} based on observations from \citet{Piotto2002} for NGC5927, NGC0104, NGC5904, and NGC1904, represented by the colored dots. Grey dots indicate the computed spectral models.
    %(from left to right): \feh\ = –1.58, –0.77, and –0.47.
    }
    \label{fig:cmdhr}
\end{figure}

{Based on the CMDs of NGC\,1904, NGC\,5904, NGC\,0104, and NGC\,5927, respectively, we computed synthetic spectra for the metallicities \feh\ = --1.60, --1.29, --0.77, and --0.47} \revise{(see Table \ref{tab-gcs} for metallicity references)}. The coverage in \teff\ and \logg\ are shown in Figure \ref{fig:cmdhr}. % illustrates the parameters of our grid for each iron abundance. 
\revise{We consider this coverage appropriate as \citet{Sakari2014} has shown that the integrated properties are not strongly sensitive to the parameter binning in the HR coverage unless the binning is very coarse. }
Basic data of the selected clusters are shown in Table \ref{tab-gcs}.

\begin{table}
\centering
\caption{Globular clusters selected to guide the atmospheric parameter coverage.  
References: (a) \citet[][]{Harris1996, Harris2010}; (b): \citet{Dotter2010}; (c): \citet{DeAngeli2005}; \revise{(d)} \citet{Carretta2009}; \revise{(e)} \citet{Usher2017}; \revise{(f)} Projected core radius in arcmin \citep[][]{Harris1996, Harris2010}. 
% \VB{[We have updated Galactic GC masses in this catalog here \citet{Hilker2020}. Perhaps we should consider it.]}
}

\begin{tabular}{ccccc}
\toprule
GC & [Fe/H] & Age & $\log M/M_{\odot}$\revise{\textsuperscript{(e)}} & $R_c$ \\
\midrule
NGC\,5927 & --\,0.49\revise{\textsuperscript{(a)}} &  12.2\textsuperscript{(b)} &            5.4 & 22  \\
NGC\,0104 & \revise{--\,0.77\textsuperscript{(d)}} &  12.8\textsuperscript{(b)} &            6.0 & 10  \\
NGC\,5904 & --\,1.29\revise{\textsuperscript{(a)}} & 12.2\textsuperscript{(b)} &            5.8 & 26  \\
NGC\,1904 & --\,1.60\revise{\textsuperscript{(a)}} & 13.0\textsuperscript{(c)} &            5.4 & 25 \\
\bottomrule
\end{tabular}
\label{tab-gcs}
\end{table}

For each \feh, we considered two chemical abundance patterns:

\begin{itemize}

\item a {standard} metal mixture with \afe\ $\sim$ 0.4 and initial He mass fraction Y = 0.256, as representative of the first generation (hereafter ``\textbf{1P}''), and;

\item a second generation whose metal composition has C decreased by 0.30 dex, N increased by 1.20 dex, O decreased by 0.45 dex, and Na increased by 0.60 dex with respect to the first generation $\alpha$-enhanced mixture, with Y = 0.300 (hereafter ``\textbf{2P}''). 
The chemical abundances of the grids are summarised in Table \ref{tab:padraoabunds}. 

\end{itemize}

\revise{These choices follow the chemical patterns adopted in C11 on the modelling of a typical metal-rich Galactic GC. They were chosen by those authors based on \citet{Carretta2005} to represent values close to the upper end of the observed anticorrelation in Galactic GCs and ensure that both populations have the same C+N+O sum to match the assumption of their adopted isochrones. }
%Our work considers four GCs with metallicities that fall within the metallicity range of \citet{Carretta2010} study ($-2.344 \leq$ \feh\ $\leq -0.441$, see Table 1), where 19 clusters exhibit CNONa anticorrelations. Nevertheless, 

\revise{To keep the assumptions of the grid homogeneous, we adopted the same abundances of CNONa and He in 2P for all the GCs in Table \ref{tab-gcs}. The chosen upper values of CNONa abundances are appropriate for a range in \feh\ from -2.0 to -0.7 dex (as of \citealt{Carretta2005}) but may be an extrapolation for our most metal-rich GC.} 

The synthetic stellar spectra are available in the POLLUX database\footnote{\url{https://pollux.oreme.org/}} and \url{http://specmodels.iag.usp.br}.

\begin{table*}
\centering

\caption{Chemical abundance patterns for each SSP.
\textbf{Y} is the normalized mass fractions of He;
%\textbf{H} (b) and \textbf{He} (c) are number fractions of H and He, respectively; \textbf{C}, \textbf{N}, \textbf{O} and \textbf{Na} are \revise{abundances as used in \atlas.}
\textbf{H} to \textbf{Na} are abundances as used in \atlas\ 
(H and He are linear number fractions, and 
C, N, O, and Na are number fractions in logarithmic scale).
}

\label{tab:padraoabunds}
\begin{tabular}{|c|c|c|c|c|c|c|c|c|c|c|}
 \toprule
 \footnotesize \textbf{Grid} &
 \footnotesize \textbf{\feh} & \footnotesize \textbf{\afe} &
 \footnotesize \textbf{Mixture} &
 \footnotesize \textbf{Y} & \footnotesize \textbf{H} & \footnotesize \textbf{He} & \footnotesize \textbf{C} & \footnotesize \textbf{N} & \footnotesize \textbf{O} & \footnotesize \textbf{Na}
\\ \toprule

   I & \footnotesize -1.58 &  \footnotesize 0.4 & 
   1P & \footnotesize 0.256 & \footnotesize 0.92015 &  \footnotesize 0.07980 &  \footnotesize -3.61 &  \footnotesize -4.21 &  \footnotesize -2.95 &  \footnotesize -5.80
 \\ %\hline
   II & \footnotesize -1.58 &  \footnotesize 0.4 & 
   2P & \footnotesize 0.300 & \footnotesize 0.90246 &  \footnotesize 0.09749 &  \footnotesize -3.91 &  \footnotesize -3.01 &  \footnotesize -3.40 &  \footnotesize -5.20

 \\ \midrule
   III & \footnotesize -1.29  &  \footnotesize 0.4 & 
   1P & \footnotesize 0.256 & \footnotesize 0.92003 &  \footnotesize 0.07987 &  \footnotesize -3.61 &  \footnotesize -4.21 &  \footnotesize -2.95 &  \footnotesize -5.80
\\ %\hline
   IV & \footnotesize -1.29  &  \footnotesize 0.4 &
   2P & 
   \footnotesize 0.300 & \footnotesize 0.90233 &  \footnotesize 0.09757 &  \footnotesize -3.91 &  \footnotesize -3.01 &  \footnotesize -3.40 &  \footnotesize -5.20
   
 \\ \midrule
   V & \footnotesize -0.77 &  \footnotesize 0.4 & 
   1P & \footnotesize 0.256 & \footnotesize 0.91951 &  \footnotesize 0.08017 &  \footnotesize -3.61 &  \footnotesize -4.21 &  \footnotesize -2.95 &  \footnotesize -5.80
\\ %\hline
   VI & \footnotesize -0.77 &  \footnotesize 0.4 & 
   2P & \footnotesize 0.300 & \footnotesize 0.90174 &  \footnotesize 0.09793 &  \footnotesize -3.91 &  \footnotesize -3.01 &  \footnotesize -3.40 &  \footnotesize -5.20
   
 \\ \midrule
   VII & \footnotesize -0.47  &  \footnotesize 0.4 & 
   1P & \footnotesize 0.256 & \footnotesize 0.91877 &  \footnotesize 0.08059 &  \footnotesize -3.61 &  \footnotesize -4.21 &  \footnotesize -2.95 &  \footnotesize -5.80
\\ %\hline
   VIII & \footnotesize -0.47  &  \footnotesize 0.4 &
   2P & 
   \footnotesize 0.300 & \footnotesize 0.90090 &  \footnotesize 0.09844 &  \footnotesize -3.91 &  \footnotesize -3.01 &  \footnotesize -3.40 &  \footnotesize -5.20

 \\ \bottomrule
\end{tabular}
\end{table*}
\section{Application to integrated spectra}
\label{sec:application}

We computed integrated spectra of stellar populations following the method proposed by M19 \citep[see as well][]{schiavon+04a, Colucci2011, Sakari2014}. In short, data from a CMD is converted to the Kiel plane (\teff\ vs \logg) via a color-\teff\ transformation. Each star in the Kiel plane is associated with a model in the stellar synthetic grid (the model that is closest to the observed star in atmospheric parameters, see equation 3 in M19). The integrated spectrum is obtained by summing up the individual model spectra, weighted by the magnitude $M_V$ of the corresponding observed stars. 

\revise{This approach has the advantage of avoiding uncertainties related to the IMF and isochrones modelling. On the other hand, it has the disadvantage of being sensitive to the sampling of luminous stars and rapid phases, incompleteness of low-mass stars, and mass segregation if present (see e.g. \citealt{McWilliamBernstein2008}), as the CMD will rarely (if ever) cover the entirety of the GC stars.
The CMDs we use in this work were observed with the WFPC2 camera with the PC centred on the cluster centre \citep{Piotto2002}. 
% The WFPC2 camera covers an area of about 2.5 x 2.5 arcmin in the sky. 
Considering the core radius of the GCs in Table \ref{tab-gcs}, our models are representative of the central parts of the GCs populations}. 

The integrated flux $F_{\lambda}$ of $N$ stars is given by:
\begin{equation}
F_\lambda=\sum_{i=1}^{N}f_{\lambda,i}C_i
\label{eq:intflux}
\end{equation}

\noindent where $f_\lambda$ is the spectrum of $i$-th star, and $C_i$ is the weight of the $i$-th star, defined as:
\begin{equation}
C_i=\frac{10^{\frac{-M_{V,i}}{2.5}}}{\int T^{V}_\lambda f^{}_{\lambda,i} d\lambda}.
\label{eq:pesoestrela}
\end{equation}
where 
$M_{V,i}$ is the absolute magnitude of the \textit{i}-th star,
% $f^{}_{\lambda,i}$ is the \textit{i}-th stellar spectrum normalized to $\int f_{\lambda,i}d\lambda = 1$, 
and $T^{V}_\lambda$ is the response function of the V-band filter.

%\lm{ I should have asked this before, but: did you normalize the total flux of each star to 1 before applying the C factor? }
%\vb{answered by email. Paula helped me correct the flux normalization issue. It does not change the results but affects our view of the integrated light between 1P and 2P. See Figure 3, which had weird total fluxes in two of the metallicities, now it seems correct.}
%\lm{ Ok!}

\revise{For each metallicity and chemical pattern we computed an integrated spectrum, resulting in eight synthetic SSPs. 
By modelling two simple populations for each iron abundance with extreme chemical pattern values (pure 1P and pure 2P), we aim to estimate an 
approximate upper level of changes in spectrophotometric features due to CNONa and He variations, 
rather than to model real GCs.}
\revise{Future work will relax these assumptions to consider varying degrees of chemical changes and 1P/2P proportions (or intermediate subpopulations)
as a function of metallicity \citep{Carretta2009}, 
mass \citep{Schiavon2013},
and radius \citep{Sakari2014}.} 

\begin{figure*}
	\includegraphics[width=\textwidth]{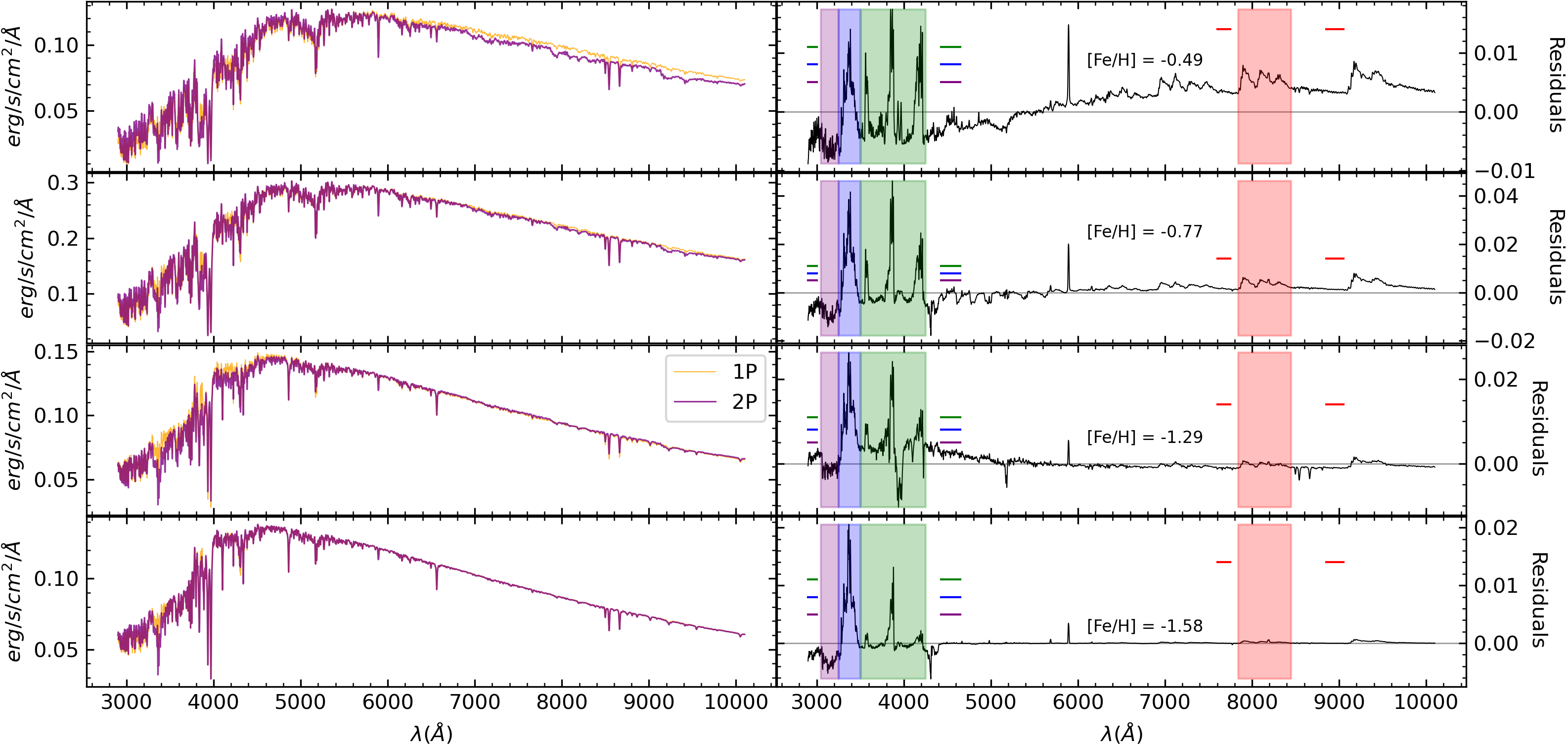}
        \caption{ 
        % Age ref: Dotter et al.(2010).	Mass ref 2010 version of the Harris(1996).	[Fe/H] ref: 2010 version of the Harris(1996). NGC1904 Age ref Mucciareli etal (2007).
        Left-side panel: The model SSPs are shown.  ``Gen.1'' indicates models with the standard mixture of the ``first generation'', i.e., "$\alpha$-enh". ``Gen.2'' represents the modified mixture of a "second generation", i.e. "$\alpha$-- and Y--enhancements and $\Delta$CNONa" (it also considers the CN--ONa abundance variations). The iron abundances are indicated in the texts on the right panels. Right-side panel: The residual flux between 1st and 2nd generation spectra, for each iron abundance. The wavelength regions affected by the chemical variations typical of the 2nd generation are seen, corresponding to changes in the CH/OH/NH and CN bands and Na D line strength. Coloured lines indicate the blue and red continua bandpass intervals that correspond to the central bandpass areas of the same colour.}
    \label{fig:specintp1p2}
\end{figure*}

The resulting integrated spectra %of each population are used as a reference for further analysis, and they 
are shown in the left-side panel in Figure \ref{fig:specintp1p2}. 
{As expected from the CMDs, the spectral energy distribution of the two most metal-poor cases shows the Balmer features caused by the strong contribution of relatively hot horizontal branch stars.} 
%\vb{[is this still observed?]}
The right-side panel of the figure shows the ratio between the 2P and 1P reference spectrum for each iron abundance. 
As can be seen from this panel, the effect of the chemical variations is stronger in the blue region, corresponding to molecules CH, OH, and NH. A feature corresponding to the \NaD\ line is in 5800\,\AA. Redder than 7000\,\AA, the signal corresponds to CN molecular band.

%\al{Can we explain why there is a difference in flux levels in cases 1 and 3, and not in cases 2 and 4?}. 
%\lm{Excellent question. To understand better I see my question on paragraph above.  }
%\pc{i proposed via email some tests to figure this out.}
%\vb{it seems this issue is fixed now. I had to do with the flux used the normalization of the weights. I was using the rectified flux in equation 3 and the total flux in equation 2 instead of using the total flux in both equations.}
%\lm{OK!}

The highlighted regions in the right-side panel in Figure \ref{fig:specintp1p2} correspond to wavelength windows with a strong sensitivity to the presence of 2P and are discussed later in Section \ref{sec:discussion_results}.

% \begin{table}
% \caption{New indices proposed to be sensitive to the 2nd generation in globular clusters. The bandpasses are defined in angstroms \AA and indices are measured in magnitudes.}
% \begin{center}
% \begin{tabular}{lccc}
% \hline
%    & Blue bandpass & Central bandpass & Red bandpass \\
% \hline
% OH blue & 2950 -- 3050 & 3050 -- 3300 & 3650 -- 3750 \\
% NH blue & 2950 -- 3050 & 3250 -- 3500 & 3650 -- 3750 \\
% CN blue & 3000 -- 3050 & 3510 -- 4250 & 4430 -- 4650 \\
% CN red  & 7845 -- 8630 & 7845 -- 8630 & 8900 -- 9050  \\
% \hline
% \end{tabular}
% \end{center}
% \label{tab:newindices}
% \end{table}%

\section{Simulating stochastic populations}
\label{sec:bootstrap}

% \revise{Globular clusters may differ, for example, by their age or mass. But, at similar physical parameters, they may have different metallicities and stellar formation histories. In that sense, clusters have different stellar contributions to the integrated spectrum: while giants contribute to the light and fluctuations, Turn-Off stars, in comparison, contribute less to the light but not to the fluctuations. This is related to the stochasticity among GC populations. Hence, we normalize our samples on the regime in which our analysis is not affected by stochastic effects, leveling them up and having comparable systems.}

% \al{The paragraph above is still confusing: there is no reason to distinguish age and mass from metallicity or star formation history in the discussion of stochastic fluctuations. I would suggest something like the following instead :} \\
{Globular clusters may differ by their global physical parameters such as age, metallicity, stellar mass, detailed star formation history, and relative fraction of 1P and 2P stars.
% \vb{ \citep{Hilker2020}}
But even clusters with the same physical parameters might have different integrated emission spectra due to the stochastic nature of the actual distribution of their finite number of stars. 
{The smaller the total mass of the cluster, the larger the relative fluctuations \revise{in observational properties, such as colours and spectroscopic indices} \citep[e.g.][]{Fouesneau2010, cervino2013}}. 
Here, we aim to examine to what extent the effects of MPs on the integrated light remain detectable in this stochastic context.} 
{To that end, we simulate stochastic populations in each of the synthetic SSPs and repeat the comparison between 1P and 2P.}

\revise{The relative amplitude of stochastic fluctuations depends on the size (total stellar mass) of the stellar population considered, and that size varies among our four Milky Way GC datasets. Most of the variance comes from 
bright and short-lived phases, such as the AGB at red or near-infrared wavelengths, or the HB at blue wavelengths, when the HB is extended.
In order to bring the four GCs to a common scale, we need to normalize the CMDs to have the same number of stars in a chosen magnitude range. 
We choose the upper limit of the luminosity range to be the Turn-Off point (TO) because 
the sampling of MS stars is stable against stochastic fluctuations.
Regarding the lower luminosity limit, we should consider that \citet{Piotto2002} observations are magnitude-limited and therefore each CMD reaches different depths in the faint MS (closer GCs reach lower masses in the MS). We chose to limit the faintest star to be considered for normalization to the star closest to $V\mathrm{(TO)} + 1$}.
Therefore we split each CMD into two sub-samples:

\begin{itemize}
\item \revise{``subsample A'' includes the stars from the TO down to 1 magnitude fainter, i.e. $V\mathrm{(TO)} \leq V_{star} \leq V\mathrm{(TO)} + 1$: this range is stable against stochastic fluctuations and is used to normalize the number of stars across the four CMDs;}
\item \revise{``subsample B'' span all stars brighter than the TO (V$_{star} < V\mathrm{(TO)}$): this is the range which causes the stochastic fluctuations in the observables (colours and/or indices).}
\end{itemize}
% which are 1 magnitude fainter than the TO up to the TO

We obtain the position of the TO (that splits subsample A and B) from the V-magnitude predicted from isochrones 
taken from ``a Bag of Stellar Tracks and Isochrones'' \citep[BaSTI; ][]{bastisoc2018, pietrinferni+21}\footnote{\url{http://basti-iac.oa-abruzzo.inaf.it/isocs.html}}.
For each GC, we requested an isochrone with the appropriate cluster age, iron abundance interpolated to that of the target cluster, $\alpha$-enhancement ($[\alpha/$Fe] = 0.4), 
and photometry in the Johnson-Cousins photometric system. 
Ages and iron abundances adopted are those in Table \ref{tab-gcs}.
For the helium mass fraction (Y), we adopted 0.247 and 0.300 for 1P and 2P, respectively.

To relate the number of stars in subsample A to the total number of stars in the cluster, we assumed that the initial stellar mass function (IMF) can be approximated with the universal function of \citet{Kroupa01}. That assumption also allows us to compute an expected number of stars in the initial mass interval associated with subsample B. The latter is (not surprisingly) not precisely equal to the actual number of stars observed in subsample B. 
We run simulations of SSPs drawing stars from the CMDs, with the constraint that
%\revise{Our bootstrapped simulations of GC stellar populations draw stars among the initial empirical sample with the constraint that} 
{every population has the same total number of stars to be integrated. 
We also set that the proportion of stars between subsamples A and B is obtained from the IMF. }

{Table \ref{fig:bootstrap} presents the quantities used to compute the number of stars. In subsample A, we use the stellar mass of the Turn-Off $M_{ini}$(TO) and the stellar mass of a star which is 1 magnitude fainter $M_{ini}$($V$(TO)\,+\,1). In subsample B, we use the stellar mass in the Turn-Off $M_{ini}$(TO) and the initial mass at the beginning of the TP-AGB phase $M_{ini}$(TP-AGB).}
\begin{table*}
\centering

\caption{
{Data obtained from the isochrones to compute the ratio between subsamples A and B.}
Reference metallicity and chemical mixtures are presented in the first two columns. 
From the third column on, we show the data obtained from the isochrones: 
% absolute V magnitude relative to the initial 
% mass $V$($M_{ini}$); 
% initial mass $M_{ini}$; 
absolute V magnitude in the Turn-Off [$V$(TO)] and 
its stellar mass [$M_{ini}$(TO)];
absolute V magnitude of the \revise{nearest neighbor star} 1 magnitude fainter than the Turn-Off [$V$(TO)\,+\,1]
and its initial mass [$M_{ini}$($V$(TO)\,+\,1)]; 
absolute V magnitude of a star at the beginning of the TP-AGB phase [$V$(TP-AGB)] and its initial mass [$M_{ini}$(TP-AGB)].}

\begin{tabular}{cc|cccccc}

\toprule

Metallicity  &      
Mixture &  
% $M_{ini}$ &  
% $V$($M_{ini}$) & 
$V$(TO) &  
$M_{ini}$(TO) &  
$V$(TO)\,+\,1 & 
$M_{ini}$($V$(TO)\,+\,1) & 
$V$(TP-AGB) &  
$M_{ini}$(TP-AGB) \\

\midrule

-0.49 &    &  4.75 & 0.879 & 5.75 &                  0.796 &     -1.08 &         0.965 
\\
-0.77 & 1P &  4.74 & 0.819 & 5.73 &                  0.742 &     -1.43 &         0.891 
\\
-1.29 &    &  4.51 & 0.773 & 5.51 &                  0.704 &     -1.86 &         0.830 
\\
-1.60 &    &  4.42 & 0.756 & 5.41 &                  0.694 &     -1.93 &         0.800 
\\
\midrule
-0.49 &    &  4.83 & 0.822 & 5.82 &                  0.744 &     -1.13 &         0.899 
\\
-0.77 &    &  4.86 & 0.753 & 5.85 &                  0.683 &     -1.47 &         0.819 
\\
-1.29 & 2P &  4.59 & 0.708 & 5.59 &                  0.646 &     -1.86 &         0.756 
\\
-1.60 &    &  4.51 & 0.692 & 5.49 &                  0.638 &     -1.88 &         0.730
\\

\bottomrule

\end{tabular}
\label{tab:vmags_masses}
\end{table*}
% \end{center}

{For each SSP in Table \ref{tab:vmags_masses} 
% of each cluster 
%accounting for the stochastic effects for all chemical hypotheses, 
we built 100 simulated populations, each {with 
20\,000 stars in the mass range between $M_{ini}(V({\mathrm{TO}})+1)$ and $M_{ini}({\mathrm{TPAGB})}$, to be representative of GCs with total masses $M\approx 10^5 M_{\odot}$}
%one composed of the same total number of stars. 
%We defined the total number of stars in each population as 20,000 stars to be representative of real GCs \revise{with $M\approx10^5 M_\odot$}. 
%\PC{i think GCs have many more stars then 20k no? this sentence need a reference or some supporting argument}
%\VB{I included the mass argument, because that's why, for intance, we are using 20k, not 10k.} \AL{OK with the above reformulation?}.
For subsample A, we bootstrapped the stars only once given that the contribution of these stars is very stable.
For subsample B, we bootstrapped the stars 100 times, one for each simulated population.
The ratio between the two subsamples is fixed, based on the IMF. 
%These populations have stars selected randomly for subsample \textit{a} (which were fixed for all simulations of each chemical mixture) and for subsample \textit{b} (which could vary amid simulations), keeping their ratio based on the IMF restrictions.}  
% \lm{[I tried to rewrite the paragraph below (now commented in the tex file) with other words to make it more clear. See if you prefer it this way and if it makes any sense.]}
%Finally, to build up each population, subsample \textit{a} has models that were randomly selected representing the stars below the TO, and subsample \textit{b} has models that were randomly selected one hundred times.
%\pc{I found this sentence confusing. I understood the whole discussion is that subsample a does not need to be bootstrapped, so why select randomly there?}.
%\vb{because the proportion of stars between the two subsample may require a number of points for subsample \textit{a} that is different from that of the empirical data in the same V(TO)--V(TO)+1 interval. So, I randomly select stars from that interval only once.}

{Therefore, each mixture and metallicity has one hundred bootstrapped populations, normalized to the same total number of stars.} Figure \ref{fig:bootstrap} illustrates the effect of the stochasticity on the \NaD\ spectral line for the two chemical mixtures of the metallicity regime relative to NGC\,0104, 47Tuc, $\mathrm{[Fe/H]} = -0.77$. Both the strength and the profile of these strong lines are sensitive to the exact sampling of the stellar mass function. The stochastic effects on the spectrophotometric indices sensitive to the MP are discussed in the next section. 

\begin{figure}
    \includegraphics[width=\columnwidth]{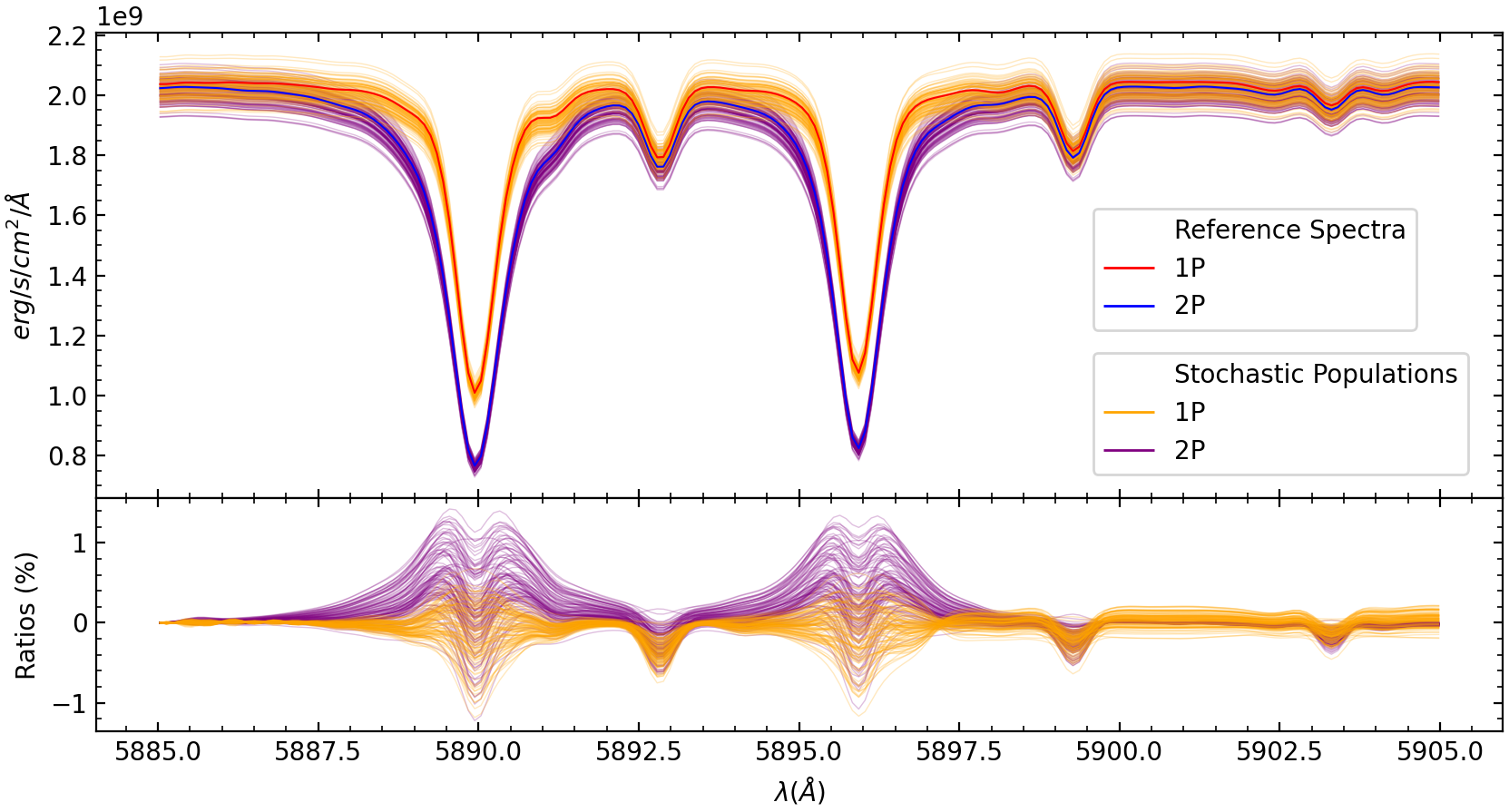}
    \caption{%\lm{Impossible to see the 1P 2P lines on top plot.... Also: were the spectra shifted for comparison?}
    %\lm{In fig. 3 this metallicity has no continuum shift, and here it seems there is.} 
    %\vb{Fixed.}
    Top: Stochastic effect in the Na D spectral line of the reference integrated spectra for 1P (blue) and 2P (red) and $\mathrm{[Fe/H]} = -0.77$. The bootstrapped populations of each mixture are shown, respectively, in orange and purple. Bottom: ratios and residuals between bootstrapped and reference integrated spectra. 
    % \pc{[I think you have not defined "the reference spectra" in the text. We should either define it or not use this expression, rather say it is the spectra without the bootstrap.]}
 %   \vb{[I have updated the text defining the "reference spectra" term, right after Equation \ref{eq:pesoestrela}, just before mentioning Figure \ref{fig:specintp1p2}.]}
    }
\label{fig:bootstrap}
\end{figure}

\section{Results and discussion}
\label{sec:discussion_results}

We evaluate the effect of abundance variations between 1P and 2P and related stochastic effects via 
spectroscopic \citep{trager+98} and %index-index diagrams 
%as in \textit{C11}, {including the new} spectral indices \revise{we define in this work}. Also, we tested the effects of the MPs with the 
photometric indices %widely used to identify MPs in resolved GCs 
\citep{Milone2016}. 
All measurements were made with the \textsc{Python} package \texttt{PYPHOT}
\footnote{\url{https://mfouesneau.github.io/pyphot/index.html}} \citep{pyphot}.

\subsection{Indices sensitive to multiple populations}

The work of C11 identifies the Lick indices that are most sensitive to the presence of a 2P, 
namely \CNa, \CNb, \Ca, \Gindex, and \NaD.
We computed these indices in our models for $\mathrm{[Fe/H]}=-0.77$ (the only metallicity explored in C11) and
compare the results in Table \ref{tab:deltaidxlick}. 

The differences between 1P and 2P from C11 and this work differ in absolute values but show the same trend.
Since the codes are the same between the two works, we attribute the difference to the opacities. 
Regarding the molecular opacities, we used \citet{CHmass2014} opacity table for the CH molecule, which is more up-to-date than the one used in C11.
Our atomic line list has been refined from literature, as explained in section \ref{calibration}.

Inspired by C11 and based on Figure \ref{fig:specintp1p2}, %in section \ref{sec:application}, 
we proposed four new spectral regions predicted to be sensitive to the CH, CN, OH, and NH molecular features in the integrated light. 
The regions were selected by visually inspecting the ratio between 1P and 2P integrated spectra, shown in the right-side panel of Figure \ref{fig:specintp1p2}.
The features are listed in Table \ref{tab:newindices}, in terms of the usual blue, central, and red bandpasses used in Lick/IDS indices \citep[e.g.][]{Worthey94}. The index names $\rm OH_{blue}$, $\rm NH_{blue}$, $\rm CN_{blue}$, and $\rm CN_{red}$ refer only to the dominant species in the central passband (see Fig. 15 in \citealt{coelho+05}), and they are represented in Figure \ref{fig:specintp1p2} by the pink, purple, green and red colours, respectively. {The NH index is a narrow-band measurement of the spectral feature also measurable with Hubble Space Telescope filters, for instance with the 3-filter index of \citet{Milone2013} now used commonly to produce so-called ``chromosome maps".}
All measured standard Lick indices, {as well as other relevant indices from the literature}, can be found in Table \ref{tab:pyphots}, and we further discuss their behaviour in the next section.

\begin{table} \centering
\caption{Variations of Lick indices for \feh\,$=-0.7$ as in C11 ($\Delta\mathrm{\textbf{\textit{I}}}_{\mathrm{Coelho}}$) and those calculated in this work ($\Delta\mathrm{\textbf{\textit{I}}}_{\mathrm{This Work}}$), between 1P and 2P.}
\label{tab:deltaidxlick}
\begin{tabular}{|c|c|c|} 
\toprule
\textbf{Index} & 
$\Delta\mathrm{\textbf{\textit{I}}}_{\mathrm{Coelho}}$ &
$\Delta\mathrm{\textbf{\textit{I}}}_{\mathrm{This Work}}$ 
\\ \midrule
\CNa &  0.084  &   0.050 \\
\CNb &  0.087  &   0.052 \\
\Ca  & -0.651 & -0.227 \\
\Gindex  & -0.572 & -0.813 \\
\NaD &  1.346  & 0.785
\\ \bottomrule
\end{tabular}
\end{table}

% \begin{table} \centering
% \caption{Variations of Lick indices for \feh\,$=-0.7$ as in \textbf{C11} ($\Delta\mathrm{\textbf{\textit{I}}}_{\mathrm{Coelho}}$) and those calculated in this work ($\Delta\mathrm{\textbf{\textit{I}}}_{\mathrm{This Work}}$), between the first and second generations.}
% \label{tab:deltaidxlick}
% \begin{tabular}{|c|c|c|} 
% \toprule
% \textbf{Index} & 
% $\Delta\mathrm{\textbf{\textit{I}}}_{\mathrm{Coelho}}$ &
% $\Delta\mathrm{\textbf{\textit{I}}}_{\mathrm{This Work}}$ 
% \\ \midrule
% CN\textsubscript{1} &  0.084  &   0.100 \\
% CN\textsubscript{2} &  0.087  &   0.103 \\
% Ca\textit{4227}  & -0.651 & -0.769 \\
% G\textit{4300}  & -0.572 & -0.584 \\
% Na\textit{D} &  1.346  & 1.033
% \\ \bottomrule
% \end{tabular}
% \end{table}

\begin{table}
    \caption{New indices proposed to be sensitive to the MP effects in GCs. The bandpasses are defined in angstroms (\AA) and indices are measured in magnitudes.}
    \begin{center}
    \begin{tabular}{lccc}
    \hline
       &  & Bandpasses &  \\
    % \hline
       & Blue  & Central  & Red  \\
    \hline
    $\rm OH_{blue}$ & 2900 -- 3000 & 3050 -- 3250 & 4430 -- 4650 \\
    $\rm NH_{blue}$ & 2900 -- 3000 & 3250 -- 3500 & 4430 -- 4650 \\
    $\rm CN_{blue}$ & 2900 -- 3000 & 3510 -- 4250 & 4430 -- 4650 \\
    $\rm CN_{red}$  & 7600 -- 7750 & 7840 -- 8450 & 8850 -- 9050  \\
    \hline
    \end{tabular}
    \end{center}
    \label{tab:newindices}
\end{table}

% \begin{figure*}
% 	\includegraphics[width=2\columnwidth]{figs/spectra/refintspecs_ratios_newidxareas.png}
%     \includegraphics[width=2\columnwidth]{figs/spectra/refintspecs_residuals_newidxareas.png}     
% 	\caption{Figura corresponde às regioes definidas dos novos indices spectrais sensiveis às MPs.}
% \label{fig:newidxareas}
% \end{figure*}      

% A Tabela \ref{tab:newindices} corresponde aos quatro indices espectrais propostos pelo nosso trabalho.
% A Tabela \ref{tab:deltaidxlick} corresponde à reprodução dos resultados obtidos em C11 para 47Tuc.

% \begin{table} \centering
% \caption{Variations of Lick indices for \feh\,$=-0.7$ as in \textbf{C11} ($\Delta\mathrm{\textbf{\textit{I}}}_{\mathrm{Coelho}}$) and those calculated in this work ($\Delta\mathrm{\textbf{\textit{I}}}_{\mathrm{This Work}}$), between the first and second generations.}
% \label{tab:deltaidxlick}
% \begin{tabular}{|c|c|c|} 
% \toprule
% \textbf{Index} & 
% $\Delta\mathrm{\textbf{\textit{I}}}_{\mathrm{Coelho}}$ &
% $\Delta\mathrm{\textbf{\textit{I}}}_{\mathrm{This Work}}$ 
% \\ \midrule
% CN\textsubscript{1} &  0.084  &   0.100 \\
% CN\textsubscript{2} &  0.087  &   0.103 \\
% Ca\textit{4227}  & -0.651 & -0.769 \\
% G\textit{4300}  & -0.572 & -0.584 \\
% Na\textit{D} &  1.346  & 1.033
% \\ \bottomrule
% \end{tabular}
% \end{table}

\subsection{The effects across different metallicities}
\label{sec:sub:effects_across_metallicities}

Figure \ref{fig:spectral_indices} shows index-index diagrams with the populations computed in this work.
Each panel shows a spectral index from those identified by C11 to be strongly sensitive to MPs
\emph{versus} the metallicity-sensitive index Fe5270. 
Each dot in the diagrams is a bootstrapped population (section \ref{sec:bootstrap}).% to contain 20,000 stars.

We confirm the results by C11 that the indices \CNa, \CNb, \Ca, \Gindex, and \NaD\ are sensitive to the presence of MPs\footnote{For conciseness we show \CNa but not \CNb, which behaves very similarly.}.
We also identify that \revise{the difference between the 1P and 2P models, represented by the line strengths in the index-index diagrams, generally increases with metallicity. This result is supported by \citet{Sakari2016, Sakari2021} where the increasing Na D line strength indicates an increasing Na-enhanced stellar abundance.}
The stochastic effect together with typical measurement errors at the two lowest metallicities prevents the separation between 1P and 2P for the indices \CNa, and \Gindex.
At higher metallicities and in the case of \NaD, and \Ca, the separation between 1P and 2P seems to be robust against stochasticity. 
In any case, one should be aware that our simulated 1P and 2P are ideal cases illustrating the maximum separations, while in integrated extragalactic GCs we will observe a mixture of populations. 

{Regarding our proposed indices, we show in Figure \ref{fig:new_spectral_indices} that some combinations of indices are robust against the stochastic effects}. 
OH-- and NH--Fe5270 diagrams do distinguish them well.
$\rm CN_{red}$ effectively split the two mixtures at high metallicity regimes, despite the high dispersion due to stochasticity in low metallicities. $\rm CN_{blue}$ has tighter splits in the high metallicities compared to its red counterpart, being not as effective as the other indices. 

In Figure \ref{fig:new_spectral_indices_mixed}, we draw attention to the OH--NH diagram (top-left panel), where the groups are well separated. 
%everything is very discretized. 
Something similar happens when comparing OH or NH with CN$_{\rm red}$ or CN$_{\rm blue}$, as in the diagrams of the second row.

We point out the work by \citet{Bertone2020newspecindices} where a set of new spectral indices was also proposed to identify the MP effects. These authors performed a star-to-star comparison at different evolutionary stages, with four chemical mixture assumptions for a metallicity $\mathrm{[Fe/H]}=-1.62$; they did not directly examine the indices expected for the integrated light of a cluster.  
Our methods for producing spectra are slightly different regarding the opacity tables and {\tt ATLAS} version ({\tt ATLAS9} there vs. {\tt ATLAS12} here). Their synthetic spectra, also computed with \synthe, cover a smaller wavelength interval at lower spectral resolutions; our OH$_{\rm blue}$ index (Table\,\ref{tab:newindices}) lies outside that interval. 
% \revise{In Figure \ref{fig:bertones_NHblue} we compare the different NH index definitions, also based on different nitrogen abundances. It is observed that all chemical mixtures are split, being as high as about 70\% in the case of NH3426 \textit{vs} $\rm NH_{blue}$.
% the closest match between the two index lists occurs between our NH$_{\rm blue}$ index and their NH3426; both can separate the two chemical mixtures via their different nitrogen abundances.}
In Appendix \ref{ap:C} we present a set of plots with \citet{Bertone2020newspecindices} spectral indices computed with the integrated spectra of our bootstrapped populations in Figure \ref{ap:fig:bertones_specindices}, as well as the measurements in Table \ref{ap:tab:pyphots_bertone}. We show that their spectral indices, designed to split individual stars of sub-populations, also split 1P and 2P of integrated populations of stars in almost every case.

%\begin{figure}
%\centering
%    \includegraphics[width=\columnwidth]{figs/spectral_indices/spectral_indices_bertone_NHblue.png}
%    \caption{\citet{Bertone2020newspecindices} NH spectral indices (y-axis) as function of our $\rm NH_{blue}$ index proposed in this work.
%    }
%\label{fig:bertones_NHblue}
%          \end{figure} 

{In Figure \ref{fig:photometry}, we show a set of diagrams with the bootstrapped populations for each metallicity and chemical mixture that splits 1P and 2P mixtures using HST filters as suggested in \cite{Milone2013}.
%from left to right,
In the left panel, the color-color diagram uses filters that enhance the separations between populations, in the MS and RGB. The middle panel is a chromosome map based on the pseudo-color index c(F275W,F336W,F410M)\footnote{Defined as $\rm (mF275W-mF336W) - (mF336W-mF410M)$.} 
that intends to maximize the separations. In this panel, instead of filter F410M, we used F438W in the combination, along with the $\rm (mF275W - mF438W)$ colour, and this change augmented the separations.
%, although it needs to be tested in more depth with the integrated light approach.
% \AL{I don't understand the "although it needs ... approach". Can we remove that or reformulate it?}
% \VB{I tried not to be too assured about the statement that "this change augmented the separations" because I did not perform a "hard test" about it.}
After exploring more combinations with the HST filters, we suggest a pseudo-colour diagram with two three-filter indices, c(F275W,F336W,438W)
% \footnote{\textit{c(F275W,F336W,438W)} $\rm =(mF275W-mF336W) - (mF336W-m438W)$.} 
vs c(F275W,F438W,F814W)
% \footnote{\textit{c(F275W,F336W,F814W)} $\rm =(mF275W-mF336W) - (mF336W-F814W)$.}
, in the right panel. This diagram splits 1P and 2P with less superposition of the bootstrapped populations \revise{thus providing} a significant opportunity for measurements of these colours for characterizing the MPs using the integrated light of GCs. 
% For instance, one needs 20\% in the precision of these filters to split 1P and 2P populations of a GC with \feh\ $=-0.77$.
}
Such photometric indices, among others, are commonly used to identify the presence of MPs in colour-magnitude diagrams of galactic GCs \citep{Gontcharov2023, Milone2023a}, and have been recently applied to the infrared region with the JWST \citep{Milone2023b, Ziliotto2023}. 
% Among the three diagrams, the chromosome-map splits the populations better. 
% \AL{which one do you mean? the 3rd diagram?}
% \VB{I forgot to update this phrase. This was true when comparing Chro-maps with CMD and color-color. I commented this part since it was mentioned before already.}

%From these discussions, we expect that in the context of integrated light, the effect of MPs could be better characterized in higher metallicity regimes.

% \lm{[What modified Nitrogen means?]}
% \vb{[the abundance is modified to account for the MP effect.]}

% ############ V-cut discussion #############
    %     \begin{table}
    %         \caption{Maximum loss (columns 2 and 3) and variation (column 4) of the light contribution for all metallicities between two integrated spectra considering all GCs stars and due to the drop of low-mass stars regarding the cut in magnitude in the CMD, respectively relative to the ``normal'' and ``V-cut'' spectra in Figure \ref{fig:refintspecs_Vcut_NGC5927}. Columns 2 and 3 refer independently to 1P and 2P comparisons. Column 4 presents the difference between the two mixtures for each metallicity of column 1.}
    %         \begin{center}
    %         \begin{tabular}{cccc}
    %         \hline
    %  [Fe/H] & Max 1P loss & Max 2P loss & $\Delta_{max}\mathrm{(1P-2P)}$ \\
    %         \hline
            
    % -0.49 & 3.2\% & 2.7\% & 0.5\% \\
    
    % -0.77 & 1.6\% & 0.9\% & 0.7\% \\
    
    % -1.29 & 3.4\% & 2.8\% & 0.6\% \\
    
    % -1.58 & 2.8\% & 2.2\% & 0.6\% \\
    
    %         \hline
    %         \end{tabular}
    %         \end{center}
    %         \label{tab:losmass_loss_variation}
    %     \end{table}
% ############ V-cut discussion #############

\begin{figure}\centering
\centering
    \includegraphics[width=\columnwidth]{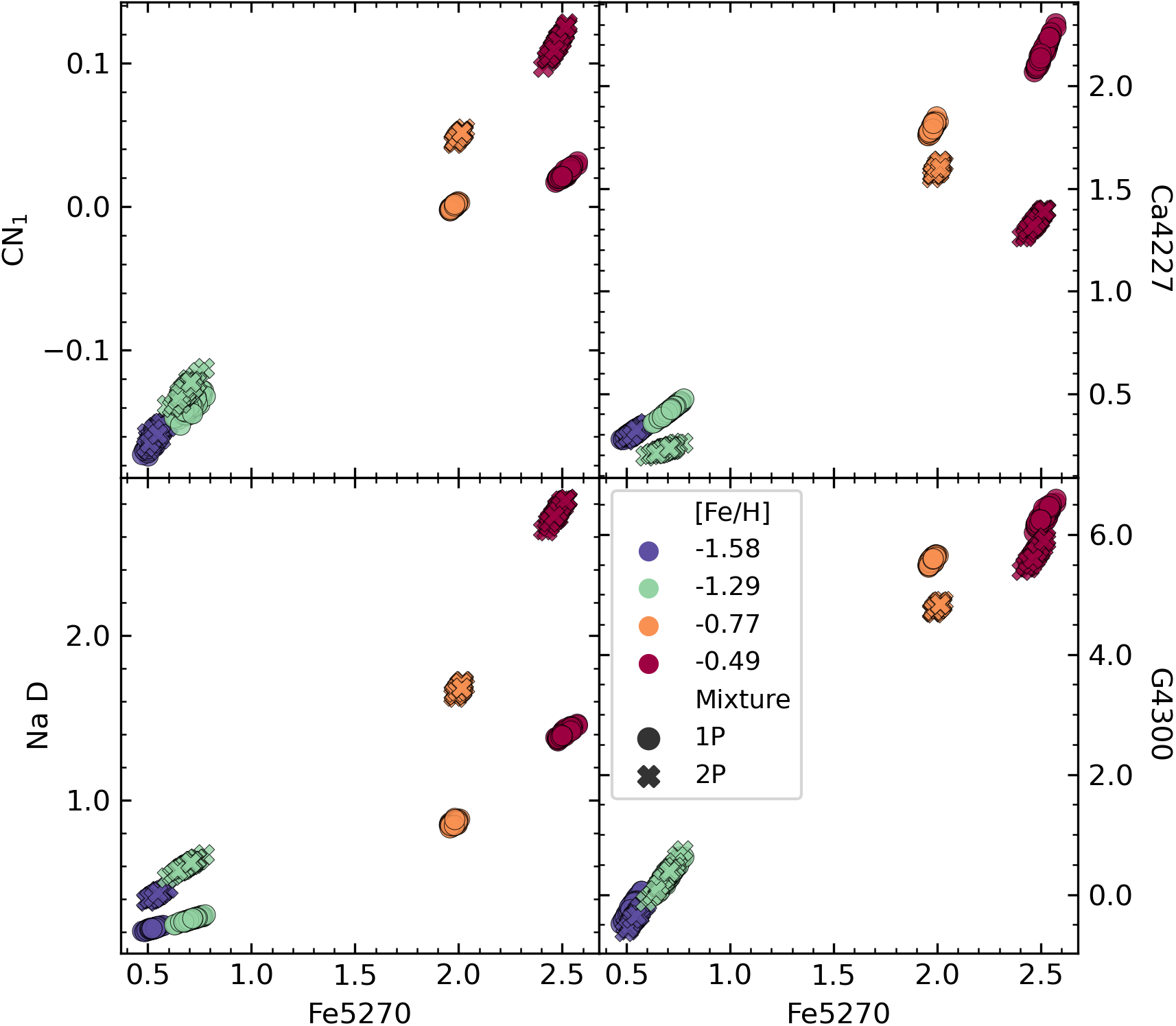}\\
\caption{The stochastic effects in the spectral indices \CNa, \Ca, \Gindex, and \NaD. Each point in the index-index diagrams corresponds to a population composed of 20,000 stellar models. Each diagram shows 100 populations computed in four metallicity regimes (coloured symbols) and two chemical mixtures (1P and 2P, rounded and x symbols, respectively).}
\label{fig:spectral_indices}
\end{figure}

\begin{figure}
\centering
\includegraphics[width=\columnwidth]{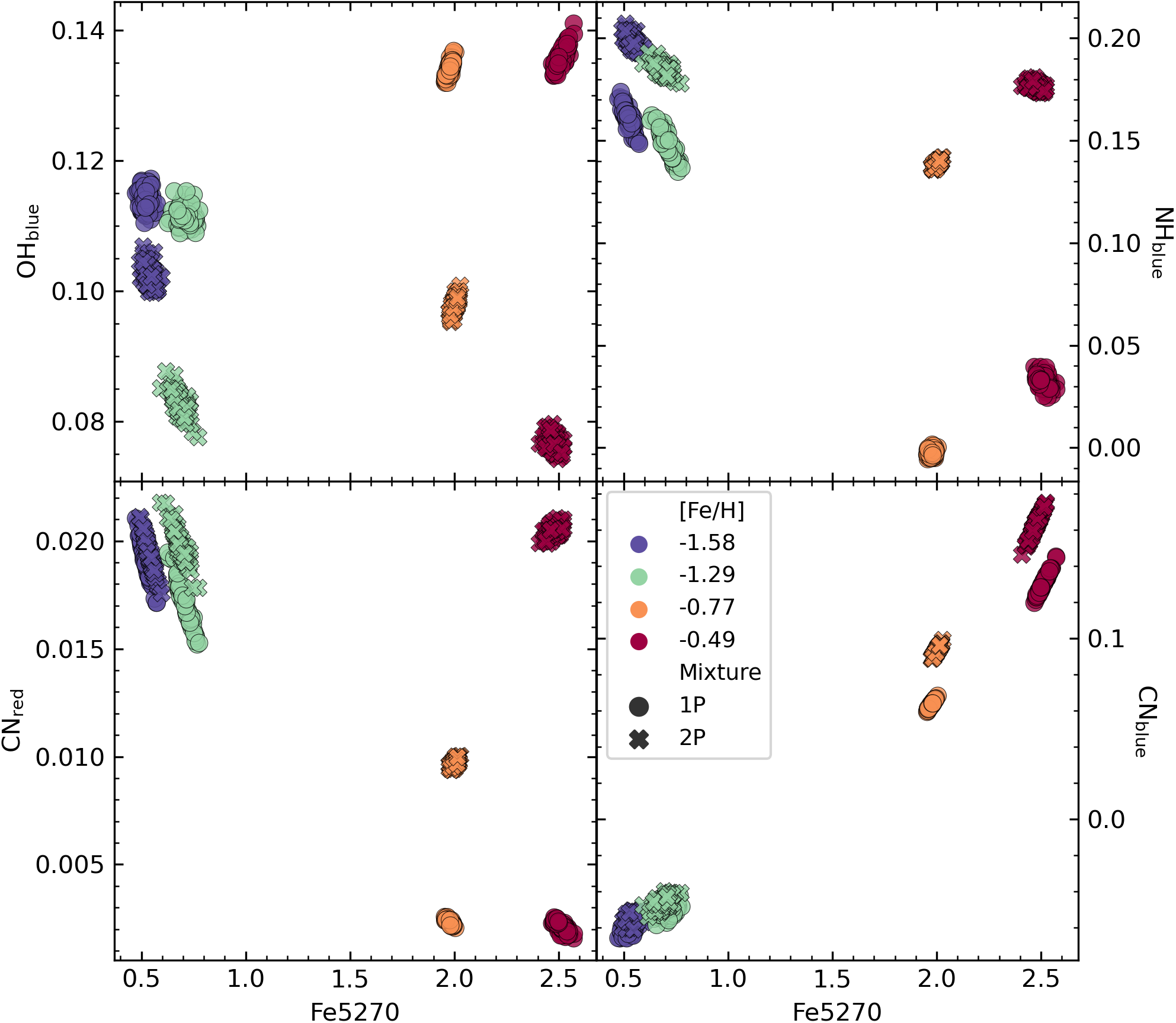}
    \caption{Same as in Figure \ref{fig:spectral_indices}; this figure presents the stochastic effects in the new indices proposed to be sensitive to the MP effects in GCs as a function of the metallicity: OH$_{\rm blue}$, NH$_{\rm blue}$, CN$_{\rm blue}$, CN$_{\rm red}$. 
    % Each point in the diagrams corresponds to a population composed of 20,000 stellar models. Each diagram shows 100 populations computed in four metallicity regimes (coloured symbols) and two chemical mixtures (``1P'' and ``2P'', rounded and \textit{x} symbols, respectively).
    }
\label{fig:new_spectral_indices}
\end{figure}

\begin{figure*}
\centering
\includegraphics[width=2\columnwidth]{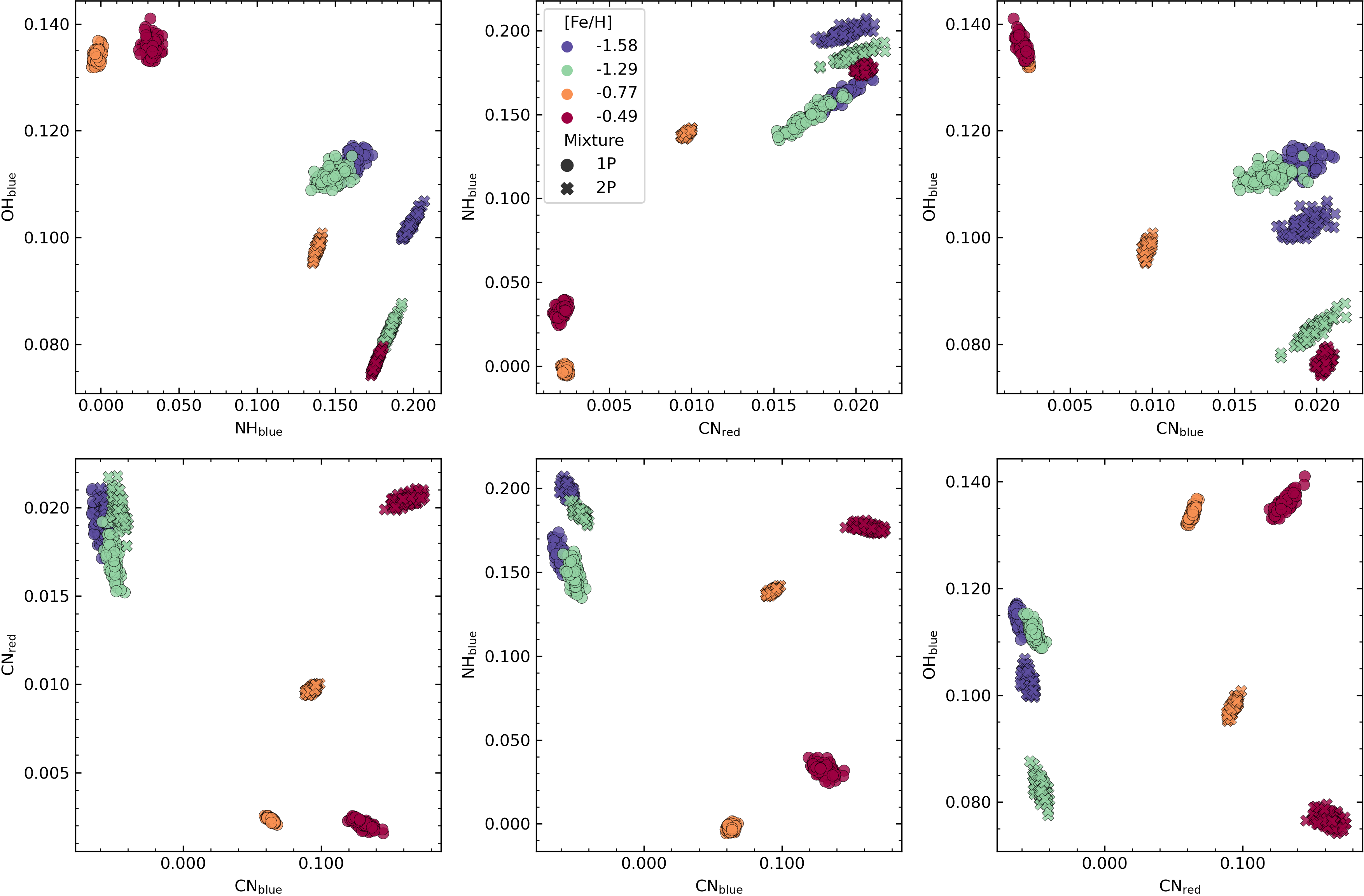}
    
    \caption{Same as in Figure \ref{fig:new_spectral_indices}; this figure presents index-index diagrams showing the stochastic effects among the new indices proposed to be sensitive to the MP effects in GCs: $\rm OH_{blue}$, $\rm NH_{blue}$, $\rm CN_{blue}$, $\rm CN_{red}$. 
    % Each point in the diagrams corresponds to a population composed of 20,000 stellar models. Each diagram shows 100 populations computed in four metallicity regimes (coloured symbols) and two chemical mixtures (``1P'' and ``2P'', rounded and \textit{x} symbols, respectively).
    }
\label{fig:new_spectral_indices_mixed}
\end{figure*}

\begin{figure*}
\centering
    \includegraphics[width=2\columnwidth]{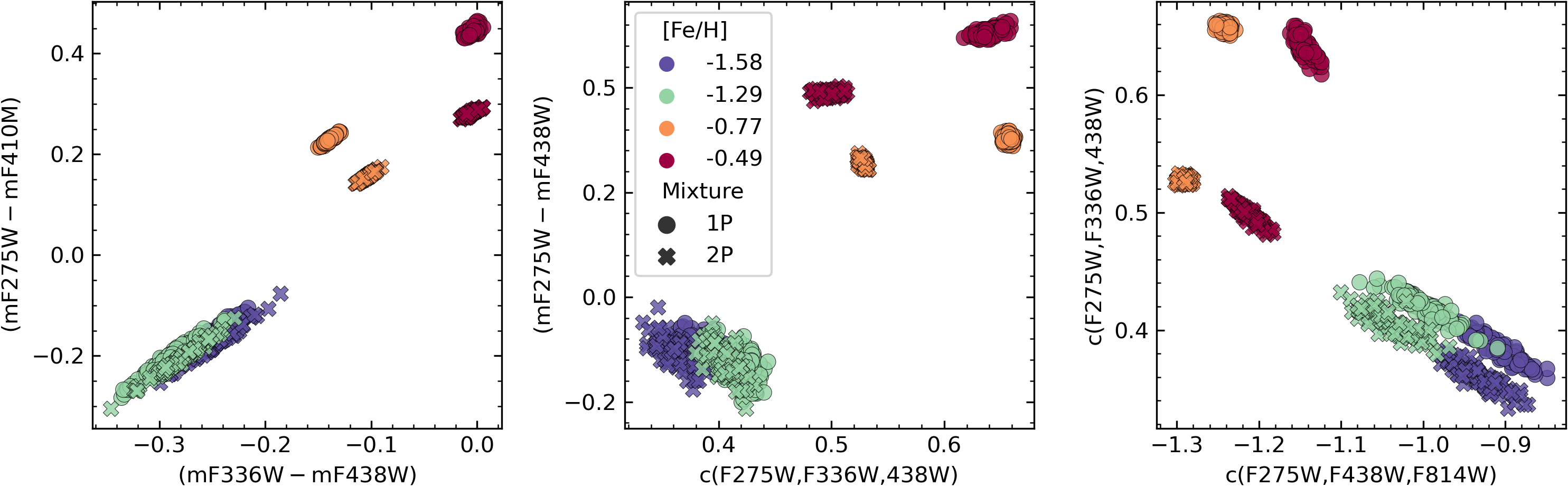}
    \caption{Same as in Figure \ref{fig:spectral_indices}; from left to right, this figure presents a colour-colour and two Chromosome-maps diagrams.
    % Each point in the photometric diagrams corresponds to a population. 
    % Each diagram shows 100 populations computed composed of 20,000 stellar models in four metallicity regimes (coloured symbols) and two chemical mixtures (``1P'' and ``2P'', rounded and \textit{x} symbols, respectively). 
    % Each point corresponds to a population composed of 20,000 stellar models.
    % \AL{======= The formula along the x-axis of the last diagram is incorrect: it should be +F438W, not -F438W. Was the actual calculation for the plot correct? I suspect not, because the results should not be numbers larger than 80. ===========}
    }
\label{fig:photometry}
\end{figure*}

\section{Conclusions}
\label{sec:conclusions}

% O que isso muda na ciência? Indique o avanço que a sua pesquisa trouxe ao cenário científico; ou seja, o que muda a partir da publicação do seu artigo.
% \\

% Neste trabalho, produzimos uma grade de espectros estelares que consideram os efeitos das anticorrelacoes quimicas.

% Também produzimos uma grade de populacoes integradas que levam em conta efeitos das MPs e de estocasticidade.

% A partir dos nossos resultados, indicamos regioes no otico que podem guiar novos trabalhos, tanto para AGs quanto outros tipos de sistemas estelares, para o estudo das MPs.

% Indices espectrais sensiveis a presenca de estrelas de baixa massa podem ser relevantes aos estudos de IMF em populacoes integradas. O efeito apareceu mais nos modelos de 1P do que 2P, variando de em 1\% para 47Tuc 

% \vb{lets remind people why the work is important}.
The multiple population phenomenon in GCs is an exciting entry point for studying these astrophysical objects. To support such studies, we publish a grid of synthetic stellar spectra \revise{at four iron abundances and two chemical abundance patterns that characterize GC stars: a standard alpha-enhanced mixture and a mixture with anticorrelated CNONa abundances and He enhancement}.

Using these spectra, we have modelled \revise{a total of eight integrated spectra of old stellar populations, a pure 1P and a pure 2P per iron abundance. The synthesis of the integrated light} is based on colour-magnitude diagrams of Milky Way globular clusters and, therefore, quantifies the effect of GC-specific \revise{abundance changes} on spectra without accounting for changes in stellar evolution that accompany big changes in the He abundance. \revise{We will address the effects of changes in isochrones in a forthcoming work.}
% But the grids of stellar spectra produced here are made available to be combined with other evolutionary assumptions in future studies.  

We \revise{quantified} the effect of different chemical mixtures on spectroscopic indices and narrow/medium photometric passbands, 
accounting for the stochastic nature of real stellar populations. 
The \revise{effect due to the presence of} 2P is detectable in many spectrophotometric indices, and the effect is stronger with increasing iron abundance.
\revise{In low-metallicities, the stochastic nature of the integrated light of GCs will hamper the detection of the presence of a second population.}

We suggest \revise{four} spectral regions that can \revise{be used} in the studies of MPs in integrated light. Our proposed spectral indices
seem to be a promising way of distinguishing populations in index-index diagrams, even in the presence of stochastic effects. 
%\revise{Given the complexity of the integrated lights, finding regions most affected by chemical changes can infer the MP effects.}
\revise{For instance, we predict that our proposed $\rm OH_{blue}$--$\rm NH_{blue}$ spectral index diagram is a strong indicator of the presence of a second population. We also investigated the behaviour of the chromosome maps when the colours are measured in the integrated light rather than in individual stars. We conclude that the second population may be detected with the following combination of colours: c(F275W, F336W, 438W) vs c(F275W, F438W, F814W).}
% Moreover, future observations could make use of the presented chromosome-map, i.e., c(F275W,F336W,438W)
% \textit{vs} c(F275W,F438W,F814W), where we consider a slightly different pseudo-colour diagram from the literature.}

\revise{By measuring spectrophotometric indices in pure 1P and 2P populations, we estimate the interval of possible values when a CNONa-anticorrelated population is present. In real-world GCs, a mixture of these populations or sub-populations will be present, and future work will be dedicated to disentangling these components in the integrated light.}

%%%%%%%%%%%%%%%% ACKNOWLEDGMENTS %%%%%%%%%%%%%%%%%%%

\begin{acknowledgements}
% thanks all collaborators for patiently adding their knowledge and reviewing every aspect of the paper.
This study was financed in part by the Coordenação de Aperfeiçoamento de Pessoal de Nível Superior – Brazil (CAPES) – Finance Code 88887.580690/2020-00 and
% and Conselho Nacional de Desenvolvimento Cient\'ifico e Tecnol\'ogico (CNPq) under grant 200928/2022-8. It was supported 
by Agence Nationale de la Recherche, France, under project POPSYCLE (ANR-19-CE31-0022).
We also acknowledge the support from 
Conselho Nacional de Desenvolvimento Cient\'ifico e Tecnol\'ogico (CNPq) under the grants 200928/2022-8, 310555/2021-3, and 307115/2021-6, and from 
Funda\c{c}\~{a}o de Amparo \`{a} Pesquisa do Estado de S\~{a}o Paulo (FAPESP) process numbers 2021/08813-7 and 2022/03703-1.
%
% FAPESP through grant 2022/03703-1 and CNPQ through grant 307115/2021-6 for finantial support.   
\end{acknowledgements}

%%%%%%%%%%%%%%%%% REFERENCES %%%%%%%%%%%%%%%%%%%%%

\bibliographystyle{aa}
\bibliography{refs} % if your bibtex file is called example.bib

%%%%%%%%%%%%%%%%% APPENDICES %%%%%%%%%%%%%%%%%%%%%

\clearpage
\appendix
\onecolumn

\section{Molecular line lists}
\label{ap:A}
% \pc{dar nomes adequados para as secoes do apendice}

Molecular species used in all synthetic spectrum computation. TiO molecule is not present in the models with \teff\,$\geq4500$\,(K) due to computation time and weak effect in the final spectrum.

% \pc{adicionar um pequeno texto intro, e referencias esse apendice em algum ponto do texto principal}

\begin{longtable}{|c|c|c|}
\caption{List of all molecular opacity species and files used in this work.}

% \\[0.5ex] \hline
% \\[-1.8ex]

\endfirsthead

\multicolumn{3}{c}{\footnotesize{{\slshape{{\tablename} \thetable{}}} - Continuação.}}
% \\[0.5ex]

\\  \toprule
    \textbf{Molécula}
&   \textbf{Diretório}
&   \textbf{Arquivo}  
\\  \midrule

% \\[0.5ex] \hline
% \\[-1.8ex]

\endhead

\multicolumn{3}{l}{{\footnotesize{Continua na próxima página\ldots}}}\\
\endfoot
\bottomrule

\endlastfoot

\toprule
    \textbf{Molecule}
&   \textbf{Web address}
&   \textbf{File}    
\\ \toprule

 AlH [A-X]
&  \url{http://kurucz.harvard.edu/molecules/alh} 
&  alhax.asc

\\ %\hline

 AlH [B-X]
&  \url{http://kurucz.harvard.edu/molecules/alh}
&  alhxx.asc

\\ %\hline

 AlO
&  \url{http://kurucz.harvard.edu/molecules/alo}
&  alopatrascu.asc

\\ %\hline

 C\textsubscript{2} [A-X]
&  \url{http://kurucz.harvard.edu/linelists/linesmol }
&  c2ax.asc

\\ %\hline

 C\textsubscript{2} [B-A]
&  \url{http://kurucz.harvard.edu/linelists/linesmol}
&  c2ba.asc

\\ %\hline

 C\textsubscript{2} [D-A]
&  \url{http://kurucz.harvard.edu/molecules/c2}
&  c2dabrookek.asc

\\ %\hline

 C\textsubscript{2} [E-A]
&  \url{http://kurucz.harvard.edu/linelists/linesmol}
&  c2ea.asc

\\ %\hline

 CaH
&  \url{http://kurucz.harvard.edu/molecules/cah}
&  cah.asc

\\ %\hline

 CaO
&  \url{http://kurucz.harvard.edu/molecules/cao}
&  
caoyurchenko.asc

\\ %\hline

 CH
&  \url{http://kurucz.harvard.edu/molecules/ch}
&  chmasseron.asc

\\ %\hline

 CN [A-X]
&  \url{http://kurucz.harvard.edu/linelists/linesmol}
&  cnax.asc

\\ %\hline

 CN [B-X]
&  \url{http://kurucz.harvard.edu/linelists/linesmol}
&  cnbx.asc

\\ %\hline

 CN [X-X]
&  \url{http://kurucz.harvard.edu/molecules/cn}
&  cnxx12brooke.asc

\\ %\hline

 CO [A-X]
&  \url{http://kurucz.harvard.edu/linelists/linesmol}
&  coax.asc

\\ %\hline

 CO [X-X]
& \url{ http://kurucz.harvard.edu/linelists/linesmol}
&  coxx.asc

\\ %\hline

 CrH [A-X]
&  \url{http://kurucz.harvard.edu/molecules/crh} 
&  crhaxbernath.asc

\\ %\hline

 FeH [F-X]
&  \url{http://kurucz.harvard.edu/molecules/feh} 
&  fehfx.asc

\\ %\hline

 H\textsubscript{2}
&  \url{http://kurucz.harvard.edu/linelists/linesmol} &  h2.asc

\\ %\hline

 MgH
&  \url{http://kurucz.harvard.edu/molecules/mgh} 
&  mgh.asc

\\ %\hline

 MgO
&  \url{http://kurucz.harvard.edu/molecules/mgo} 
&  mgodaily.asc

\\ %\hline

 NaH
&  \url{http://kurucz.harvard.edu/molecules/nah} 
&  nahrivlin.asc

\\ %\hline

 NH
&  \url{http://kurucz.harvard.edu/linelists/linesmol} 
&  nh.asc

\\ %\hline

 OH
&  \url{http://kurucz.harvard.edu/molecules/oh}
&  ohupdate.asc

\\ %\hline

 SiH
&  \url{http://kurucz.harvard.edu/linelists/linesmol} 
&  sihnew.asc

\\ %\hline

 SiO [A-X]
&  \url{http://kurucz.harvard.edu/linelists/linesmol} 
&  sioax.asc

\\ %\hline

 SiO [E-X]
&  \url{http://kurucz.harvard.edu/linelists/linesmol} 
&  sioex.asc

\\ %\hline

 SiO [X-X]
&  \url{http://kurucz.harvard.edu/linelists/linesmol} 
&  sioxx.asc

\\ %\hline

 TiH
&  \url{http://kurucz.harvard.edu/molecules/tih} 
&  tih.asc

\\ %\hline

 TiO
&  \url{http://kurucz.harvard.edu/molecules/tio} 
&  tioschwenke.asc

\\ %\hline

 VO
&  \url{http://kurucz.harvard.edu/molecules/vo} &  vo.asc

\label{ap:tab:molecules}
\end{longtable}

\newpage
\section{Spectral indices and photometry}
\label{ap:B}

All measurements in Table \ref{tab:pyphots} refer to the chemical mixtures and metallicities of the reference integrated spectra.

\begin{longtable}{|c|c|c|c|c|c|c|c|c|}
\caption{Spectral indices and colours of each reference integrated spectrum.}

\endfirsthead

\multicolumn{9}{c}{\footnotesize{{\slshape{{\tablename} \thetable{}}} - Continuação.}}

\\  \toprule
    \textbf{[Fe/H]} & \textbf{-1.58} & \textbf{-1.58} & \textbf{-1.29} & \textbf{-1.29} & \textbf{-0.77} & \textbf{-0.77} & \textbf{-0.49 } & \textbf{-0.49 }  
\\  \midrule

\endhead

\multicolumn{9}{c}{{\footnotesize{Continue in the next page.\ldots}}}\\
\endfoot
\bottomrule

\endlastfoot

\toprule
    \textbf{[Fe/H]}   & \textbf{-0.49 } & \textbf{-0.49 } & \textbf{-0.77} & \textbf{-0.77} & \textbf{-1.29} & \textbf{-1.29} & \textbf{-1.58} & \textbf{-1.58}   
\\ \toprule

\textbf{Mixture} & 1P & 2P & 1P & 2P & 1P & 2P & 1P & 2P  \\ \hline

        \midrule

\CNa            &  0.061 &  0.191 &  0.026 &  0.098 & -0.037 & -0.009 & -0.049 & -0.037 \\
\CNb            &  0.107 &  0.239 &  0.065 &  0.139 &  0.005 &  0.030 & -0.010 &  0.003 \\
\Ca             &  2.802 &  1.776 &  2.223 &  1.884 &  1.099 &  0.426 &  0.828 &  0.758 \\
\Gindex         &  6.933 &  6.528 &  6.397 &  5.630 &  3.770 &  3.831 &  3.458 &  2.886 \\
\NaD            &  1.724 &  3.311 &  1.094 &  2.025 &  0.492 &  1.066 &  0.403 &  0.758 \\
Fe5270          &  2.849 &  2.810 &  2.254 &  2.276 &  1.350 &  1.352 &  1.098 &  1.112 \\
% Fe5335          &  0.844 &  0.854 &  1.057 &  1.124 &  1.814 &  1.855 &  2.475 &  2.361 \\
% H$\beta$        &  2.227 &  2.233 &  2.237 &  2.090 &  1.177 &  1.210 &  0.872 &  1.011 \\
% H$\delta_F$     &  2.575 &  2.547 &  2.430 &  1.967 &  0.562 &  0.362 &  0.013 & -0.502 \\
% H$\gamma_F$     &  1.352 &  1.596 &  1.158 &  0.943 & -1.504 & -1.057 & -2.537 & -2.190 \\
$\rm OH_{blue}$ &  0.179 &  0.106 &  0.169 &  0.128 &  0.126 &  0.079 &  0.119 &  0.092 \\
$\rm NH_{blue}$ &  0.081 &  0.214 &  0.037 &  0.175 &  0.052 &  0.178 &  0.046 &  0.148 \\
$\rm CN_{blue}$ &  0.211 &  0.246 &  0.117 &  0.153 &  0.013 &  0.040 & -0.008 &  0.005 \\
$\rm CN_{blue}$ &  0.000 &  0.025 &  0.000 &  0.012 &  0.004 &  0.013 &  0.004 &  0.007 \\
% Mg b            &  1.747 &  1.761 &  2.186 &  1.444 &  3.562 &  3.628 &  4.148 &  3.901 \\
$\rm [MgFe]$    &  0.471 &  0.477 &  0.565 &  0.406 &  0.885 &  0.874 &  0.881 &  0.935 \\

        \midrule

        (\footnotesize $\rm mF336W - mF438W$) &
        -0.251 & -0.242 & 
        -0.289 & -0.276 & 
        -0.124 & -0.092 & 
        0.004 & 0.003  \\ 
        
        (\footnotesize $\rm mF438W - mF814W$) & 
        0.780 & 0.777 & 
        0.829 & 0.876 & 
        1.640 & 1.626 & 
        1.786 & 1.707  \\ 
        
        \footnotesize c(F275W,F336W,F438W) & 
        0.376 & 0.347 & 
        0.412 & 0.397 & 
        0.651 & 0.522 & 
        0.636 & 0.494

\label{tab:pyphots}
\end{longtable}

\newpage
\section{Spectral indices proposed in the literature}
\label{ap:C}

\citet{Bertone2020newspecindices} spectral indices were built upon stellar spectra. Therefore, we show in Figure \ref{ap:fig:bertones_specindices} how their proposed indices split the sub-populations when applied to the integrated light of a synthetic cluster. Table \ref{ap:tab:pyphots_bertone} shows measurements of the reference integrated spectra.

\begin{figure*}[!h]
\centering
    \includegraphics[width=\columnwidth]{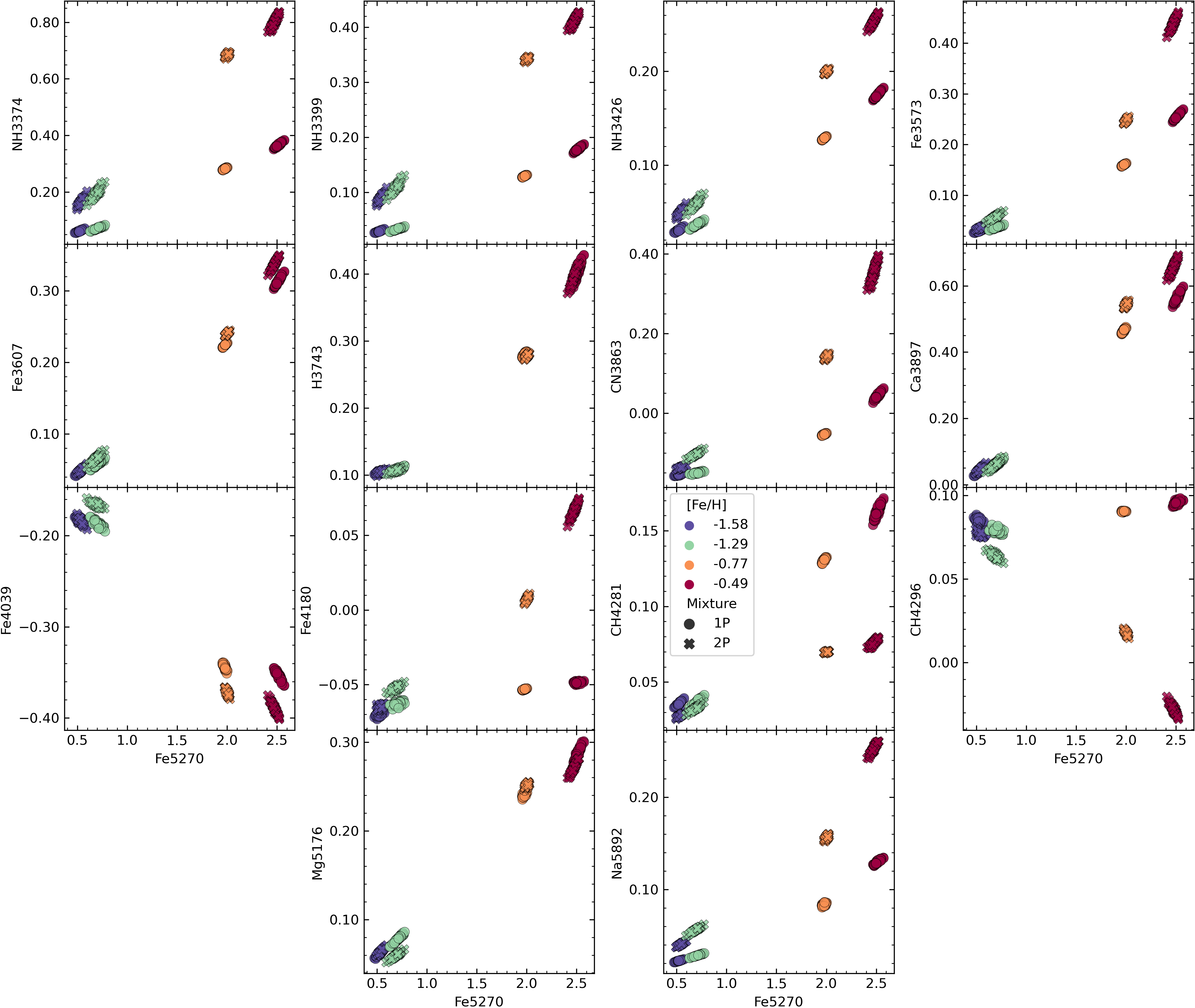}
    \caption{Diagrams with proposed spectral indices from the literature computed in the integrated light of different chemical mixture assumptions.
    }
\label{ap:fig:bertones_specindices}
\end{figure*}    

\newpage
\begin{longtable}{|c|c|c|c|c|c|c|c|c|}
\caption{Bertone spectral indices measured with our reference integrated spectra.}

\endfirsthead

\multicolumn{9}{c}{\footnotesize{{\slshape{{\tablename} \thetable{}}} - Continuation.}}

\\  \toprule
    \textbf{[Fe/H]} & \textbf{-1.58} & \textbf{-1.58} & \textbf{-1.29} & \textbf{-1.29} & \textbf{-0.77} & \textbf{-0.77} & \textbf{-0.49 } & \textbf{-0.49 }  
\\  \midrule

\endhead

\multicolumn{9}{c}{{\footnotesize{Continue in the next page.\ldots}}}\\
\endfoot
\bottomrule

\endlastfoot

\toprule
    \textbf{[Fe/H]} & \textbf{-0.49} & \textbf{-0.49} & \textbf{-0.77} & \textbf{-0.77} & \textbf{-1.29} & \textbf{-1.29} & \textbf{-1.58 } & \textbf{-1.58 }    
\\ \toprule
							
\textbf{Mixture} & 1P & 2P & 1P & 2P & 1P & 2P & 1P & 2P  \\ \hline \midrule

NH3374   &   0.445 &   0.898 &   0.350 &   0.765 &   0.192 &   0.511 &   0.172 &   0.449 \\
NH3399   &   0.217 &   0.456 &   0.161 &   0.378 &   0.087 &   0.272 &   0.078 &   0.232 \\
NH3426   &   0.218 &   0.304 &   0.159 &   0.234 &   0.087 &   0.146 &   0.075 &   0.125 \\
Fe3573   &   0.329 &   0.581 &   0.203 &   0.335 &   0.090 &   0.167 &   0.071 &   0.096 \\
Fe3607   &   0.381 &   0.415 &   0.269 &   0.289 &   0.136 &   0.156 &   0.110 &   0.116 \\
H3743    &   0.513 &   0.506 &   0.341 &   0.330 &   0.170 &   0.168 &   0.140 &   0.139 \\
CN3863   &   0.115 &   0.491 &  -0.010 &   0.239 &  -0.104 &   0.061 &  -0.106 &  -0.028 \\
Ca3897   &   0.695 &   0.810 &   0.567 &   0.662 &   0.272 &   0.282 &   0.221 &   0.250 \\
Fe4039   &  -0.413 &  -0.472 &  -0.396 &  -0.436 &  -0.273 &  -0.228 &  -0.244 &  -0.250 \\
Fe4180   &  -0.031 &   0.126 &  -0.046 &   0.043 &  -0.047 &  -0.009 &  -0.048 &  -0.030 \\
CH4281   &   0.181 &   0.077 &   0.153 &   0.075 &   0.089 &   0.072 &   0.086 &   0.058 \\
CH4296   &   0.087 &  -0.069 &   0.093 &  -0.009 &   0.074 &   0.027 &   0.073 &   0.046 \\
Mg5176   &   0.335 &   0.316 &   0.269 &   0.273 &   0.150 &   0.109 &   0.119 &   0.120 \\
Na5892   &   0.157 &   0.305 &   0.104 &   0.187 &   0.049 &   0.098 &   0.040 &   0.071

\label{ap:tab:pyphots_bertone}
\end{longtable}

\end{document}